 \documentclass[final,1p,times]{elsarticle}

\usepackage{lineno}
\usepackage{amssymb, amsfonts}
\usepackage{lipsum}
\usepackage{rotating}
\usepackage{pslatex}
\usepackage{epsfig,soul}
\usepackage[normalem]{ulem}
\usepackage{caption,subcaption}
\usepackage{booktabs,multirow}
\usepackage{graphicx,tikz,graphics,epsfig}
\usepackage{url,CJKutf8}
\usepackage{color,xcolor}
\usepackage{booktabs}
\usepackage{multirow}
\usepackage{setspace} 
\usepackage{physics,amsmath} 
\usepackage{bm}  
\journal{Computer Physics Communications}
 
\begin{document}  
\begin{frontmatter}  

 \title{Artificial and Eddy Viscosity in Large Eddy Simulation Part 2: Turbulence Models}
  \author[jsun]{Jing Sun\corref{cor1}}
 \ead{j.sun@rug.nl}  
 \author[jsun]{Roel Verstappen}
 \ead{r.w.c.p.verstappen@rug.nl}  
 \cortext[cor1]{Corresponding author}
 \affiliation[jsun]{organization={University of Groningen, Computational and Numerical Mathematics — Bernoulli Institute}, 
            addressline={Nijenborgh 9, 9747 AG, Groningen}, 
            city={Groningen}, 
            country={The Netherlands}}     
    
\begin{abstract}
This is the second part to our companion paper. The novel method to quantify artificial dissipation proposed in Part 1 
is further applied in turbulent channel flow at $\mathrm{Re_\tau}=180$ using various subgrid-scale models, with an emphasis on minimum-dissipation models (QR and AMD). We found that the amount of artificial viscosity is comparable to the eddy viscosity, but their distributions are essentially opposite. The artificial viscosity can either produce turbulent kinetic energy (TKE) in the near wall region or dissipate TKE in the bulk region. The eddy viscosity is almost uniformly distributed across the wall-normal direction and actively damps TKE at the channel center. An eddy viscosity level of $\nu_e/\nu < 22\%$ leads to accurate predictions of flow quantities at the current mesh resolution. The optimal coefficients for the QR model combined with symmetry-preserving discretization were found to be C=0.092 for turbulent kinetic energy and C=0.012 for Reynolds stress terms and large scale motions. Adjusting the QR model coefficient changes the artificial viscosity and significantly impact turbulence characteristics: small-scale motions remain unaffected, while larger scales are less accurately captured with larger coefficients. For error quantification, the SGS activity parameter $\nu_{rel} = \frac{\nu_e}{\nu_e + \nu}$ helps relate and reflect the error magnitude with the SGS model. Optimal coefficients for the AMD model were also identified. The trends of varying the coefficient are similar to the QR model but differ in the amplitude of changes. The AMD model is more sensitive to the model coefficients and introduces more eddy viscosity in the near-wall region compared to the QR model. This research highlights the importance of balancing numerical and eddy viscosities for accurate LES predictions.
\end{abstract}
\begin{keyword}
Artificial dissipation\sep  Eddy viscosity\sep Minimum-dissipation model\sep Symmetry-preserving discretization\sep Turbulence\sep Large eddy simulation
\end{keyword}  
\end{frontmatter} 

\section{Introduction}   
This is the second in a two-part series of papers, in which we quantify artificial dissipation based on the residual of the transport equation of the turbulent kinetic energy. In 
Part 1 \cite{sun2025artificialP1}, we applied our method to study the impact of the temporal and spatial schemes. This chapter is devoted to subgrid-scale (SGS) models.

Among the existing subgrid models, the Smagorinsky model is the most commonly used \cite{smagorinsky1963general}. Although it performs well in decaying homogeneous isotropic turbulence simulations \cite{lilly1966application, mansour1979improved}, it inappropriately dissipates eddies in laminar and transitional flows. The Smagorinsky model can be improved by computing the model coefficient dynamically \cite{germano1991}, but then the coefficient is to be averaged out because otherwise it will vary too much. Alternatively, the wall-adapting local eddy-viscosity (WALE) model \cite{nicoud1999subgrid} corrects behavior near walls using the square of the velocity gradient tensor. Finally, the Vreman`s model \cite{vreman2004eddy} is exact for pure shear flow.

Minimum-dissipation models offer a alternative for parameterizing subfilter turbulent fluxes in LES. The first minimum-dissipation model, the QR model, was proposed by Verstappen \cite{verstappen2011does, verstappen2018much}. The QR model is cost-effective, appropriately switches off in laminar and transitional flows, and is consistent with the exact subfilter stress tensor on isotropic grids. Rozema et al. \cite{rozema2015minimum} subsequently developed the anisotropic minimum-dissipation model (AMD) for flows on anisotropic grids. Various studies have employed the AMD model, such as those by Abkar and Moin \cite{abkar2016minimum}, Zahiri et al. \cite{zahiri2019anisotropic}, Lasota et al. \cite{lasota2023anisotropic}, Catherine et al. \cite{catherine2018} and so on. 

The model coefficient used in most of these studies aligns with the findings of Rozema et al. \cite{Rozema2020LowDissipationSM}, although a coefficient of $C = 0.0208$  was used in \cite{zahiri2019anisotropic}. Using the analytical value of the AMD model coefficient does not yield the best predictions as the effects of the numerics cannot be neglected. Artificial dissipation $\epsilon^{art}$, for instance, damps the kinetic energy, just like the eddy dissipation $\epsilon^{sgs}$. Therefore, the total dissipation, given by $\epsilon^{tot} = \epsilon^{art} +\epsilon^{sgs}$, is to be considered to obtain correct flow predictions.  

The QR model uses the second $q(v)$ and third $r(v)$ invariant of the strain-rate tensor. The second invariant corresponds to damping. The third invariant $r(v)$ measures vortex stretching. The QR model can be interpreted as follows: the eddy viscosity is taken such that the corresponding damping of the enstrophy equals the production by means of the vortex stretching mechanism. Therefore, the QR model is a promising candidate for certain scenarios, including simulations on isotropic meshes.

To obtain a numerically stable solution of Navier-Stokes equation, one needs to employ a numerical method that conserves mass, momentum, and kinetic energy in a discrete sense \cite{KRAVCHENKO1997310}. Verstappen and Veldman \cite{verstappen2003symmetry} proposed preserving the symmetry properties of underlying differential operators on unstructured staggered grids. Trias et al. \cite{TRIAS2014246} generalized this method for unstructured collocated meshes, proposing a conservative regularization of the convective term to mitigate spurious modes. Komen et al. \cite{komen2021symmetry} developed a second-order time-accurate PISO-based pressure-velocity coupling method for incompressible Navier-Stokes equations on collocated grids, implemented in OpenFOAM. The code used in this study is provided by Hopman \cite{janneshopmanRKSymFoam}.

In this work, we study the interaction between artificial viscosity and eddy viscosity under the variation of the SGS models in turbulent channel flow. The novel contributions of this paper include: (1) the quantification method for the artificial dissipation rate and the interaction between artificial viscosity and eddy viscosity in LES, (2) the optimal model coefficients for the QR and AMD models, (3) proposing and testing various static and dynamic minimum-dissipation models, (4) the impact of eddy dissipation and artificial dissipation on turbulent kinetic energy transport and turbulent characteristics, (5) the subgrid activity parameter $\nu_{rel}=\nu_e/(\nu_e +\nu)$ as an error quantification metric. 

The structure of this paper is as follows: Section \ref{sec:problem} describes the test case and propose the subgrid activity parameter. Section \ref{sec:res} presents results and discussion of turbulent channel flow simulations, validating computational results against detailed DNS data from various studies \cite{moser1999direct, hoyas2006scaling, hoyas2008reynolds, jimenez2008turbulent, juan2001direct, juan2003spectra, juan2004scaling}. We explore the consistency between isotropic and anisotropic minimum-dissipation models, optimize the QR and AMD model coefficients, and compare them to other SGS models. Section \ref{sec:sum} summarizes the conclusions.

\section{Turbulent Channel Flow Simulations} \label{sec:problem}
Turbulent channel flow is one of the most fundamental wall-bounded shear flows and it has been widely used to study the structure of near-wall turbulence \cite{moser1999direct}. In this section, numerical results of channel flow simulations using the minimum--dissipation model are presented, for friction Reynolds numbers $\mathrm{Re}_\tau =180$ (based on the half channel width). The results at Reynolds number $\mathrm{Re}_\tau= \frac{\delta u_\tau}{\nu}\approx 180$ are compared to the DNS data from Vreman \cite{vreman2014comparison}. The details regarding the spatial and temporal discretization schemes, the solver used, methods for quantifying artificial dissipation and viscosity, error quantification, relevant physical parameters, and the computational domain are described in Part 1 of this series.
 
\paragraph{Metric used to Measure Eddy Viscosity}
The relative amount of eddy viscosity is quantified using
\begin{equation}
\nu_{rel} = \frac{\langle \nu_e\rangle}{\langle \nu_e\rangle + \langle \nu \rangle }.
\label{eq:eddyToMol}
\end{equation}
This metric differs from the SGS activity parameter $s = \frac{\langle 2\nu_e \bar S_{ij} \bar S_{ij}\rangle}{\langle 2\nu_e \bar S_{ij} \bar S_{ij}\rangle + \langle 2\nu \bar S_{ij} \bar S_{ij}\rangle}$ proposed by Geurts \cite{geurts2002}. Here the SGS parameter $\langle 2\nu_e \bar S_{ij} \bar S_{ij}\rangle$ is replaced by $\nu_e$ because, for a tensor SGS model, $\langle 2\nu_e \bar S_{ij} \bar S_{ij}\rangle$ represents the projection of $2\nu_e\bar S_{ij}$ onto the direction of the symmetric part of the velocity gradient tensor $\bar S_{ij}$. The alignment between the actual subgrid stress and $\bar S_{ij}$ is unknown, so this projection may not accurately represent the error in subgrid dissipation. Therefore, the metric $\nu_{rel}$ is preferred.
 
\paragraph{Comparison of Reynolds Stress}
Since the QR model is traceless, only the deviatoric Reynolds stresses is reconstructed and directly compared with DNS and experimental data \cite{winckelmans2002comparison}. The comparison is carried out via 
\begin{equation}
    R_{ij}^{DNS,dev} \approx   R_{ij}^{LES,dev} + \langle \tau_{ij}^{SGS,dev} \rangle,
\end{equation}
where the $\langle \tau_{ij}^{SGS,dev} \rangle$ is the averaged deviatoric SGS tensor and $R_{ij}^{dev} $ is the deviatoric Reynolds stress tensor. Here, the Reynolds stress tensor is defined as 
\begin{equation*}
    R_{ij}=\langle u_i u_j \rangle - \langle u_i \rangle \langle u_i \rangle = \langle u_i'u_j' \rangle,
\end{equation*}
where $u_i$ represents the velocity. Moreover, the trace of the Reynolds stress tensor can partly be reconstructed using the turbulent kinetic energy.
\begin{align}
    R_{ij}^{DNS} &\approx   R_{ij}^{LES} + (\langle \tau_{ij}^{SGS,dev} \rangle + \frac{2}{3} \langle \bar k_{sgs} \rangle \delta_{ij})\\
    &\approx  R_{ij}^{LES} + (-2\nu_{e} S(v) + \frac{2}{3} \langle \bar k_{sgs} \rangle \delta_{ij}).
\end{align} 
\paragraph{Optimize Simulation}
There has long been debated about the intertwining of numerical and subgrid modeling errors,  in particular when an implicit LES filter is used. In this study, we aim to examine the performance of the subgrid model in a reliable and robust way. To achieve this, we conducted a sensitivity analysis on the CFL number, initial conditions, and statistical averaging time, minimizing the artificial dissipation of the precursor simulation (without a SGS model) to a level below $\epsilon^{art}_k<0.15\%$ (or $\epsilon^{art}_k/\epsilon^{\nu}_k\rightarrow 0$)  (given the current mesh resolution). Therefore, we can ensure the dominant mechanism is the subgrid model when the SGS model is add in later real simulations. 
The upcoming simulation results will include both LES simulations without subgrid model and those with subgrid models. This presentation will offer insight into the extent to which artificial dissipation contributes to the overall results. 
 
\section{Results and Discussion} \label{sec:res}
\subsection{Minimum-Dissipation Models}
Before employing SGS models for LES, it is essential to ensure their correct implementation. This study evaluates the QR, AMD, QR Grad, and AMD-Symm models on isotropic grids to verify this. Analytically, the AMD-Symm and QR models should yield identical results on isotropic grids if the model coefficients are identical. However, since the numerator of the AMD-Symm model employs the strain-rate tensor directly, while the QR model uses the determinant of the strain-rate tensor, these two models are not expected to produce identical results. The eddy viscosity expressions for these models are as follows: 
\begin{alignat}{2}
    &\nu_e = C \frac{max\{(-\Lambda \cdot \nabla v)^T (\Lambda \cdot \nabla v):S, \hspace{0.1em}0\}}{\nabla v : \nabla v},  \hspace{1.8em}&&\mathrm{AMD} \label{eq:AMD}\\
    &\nu_e = C \frac{max\{(-\Lambda \cdot S)^T (\Lambda \cdot S):S, \hspace{0.1em}0\}}{S : S},  \quad \quad \quad &&\mathrm{AMD-Symm} \\
    &\nu_e = C \delta^2 \frac{max\{(\nabla v)^T (\nabla v):S, \hspace{0.1em}0\}}{\nabla v : \nabla v},  \hspace{4.3em}&&\mathrm{QR-Grad} \\
    &\nu_e = C \delta^2 \frac{max\{r(v), \hspace{0.1em}0\}}{q(v)}, \hspace{8.9em}&&\mathrm{QR} 
\end{alignat}
where $\Lambda$ represents a diagonal matrix with its diagonal elements containing the grid widths. $(\hspace{0.2em})^T$ indicates the transpose of a tensor. $S$ for the abbreviation of $S(v)$. $\delta$ is the geometric mean of the grid size. The AMD-Symm model above is obtained by taking inner product of the grid width directly with the strain-rate-tensor, i.e., $\Lambda\cdot S$, another way to obtain this model is calculating the scaled strain-rate-tensor $\tilde S$ by taking the inner product of the grid width matrix with the velocity gradient, i.e., $\tilde S = \frac{1}{2} \left(\Lambda \cdot \nabla v + (\Lambda \cdot \nabla v)^T \right) =\frac{1}{2} \left(\Lambda \cdot \nabla v + (\nabla v)^T \cdot \Lambda^T \right)  $, resulting in the eddy viscosity
\begin{equation}
 \nu_e = C \frac{max\{(-\tilde S)^T (\tilde S):S,\,0\}}{S : S},  \quad \mathrm{AMD-Symm2}. 
\end{equation}
\label{sec:isotropic}
\begin{figure}[b!]
	\centering
	\includegraphics[width=0.32\linewidth]{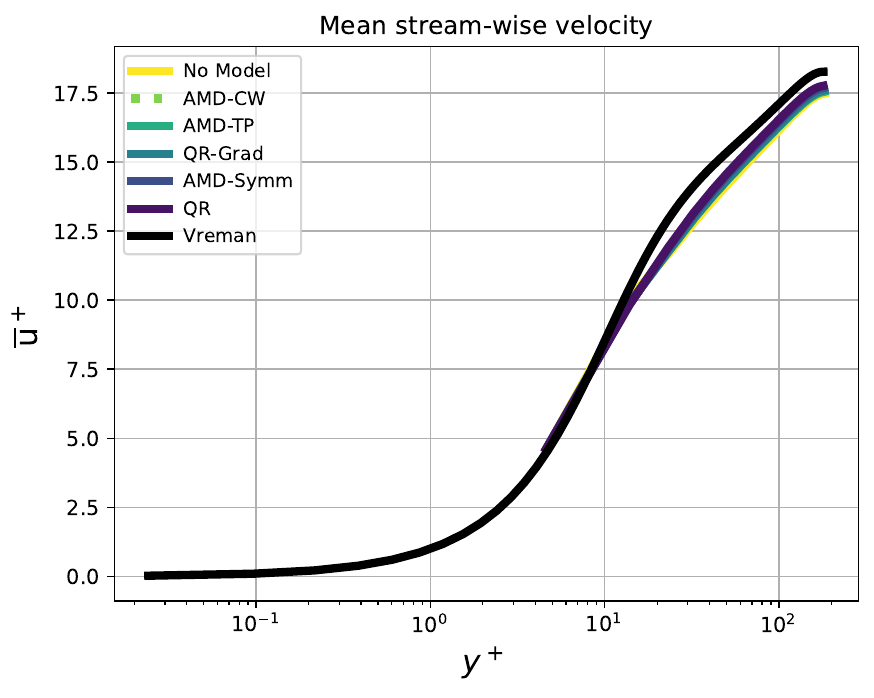}
	\includegraphics[width=0.32\linewidth]{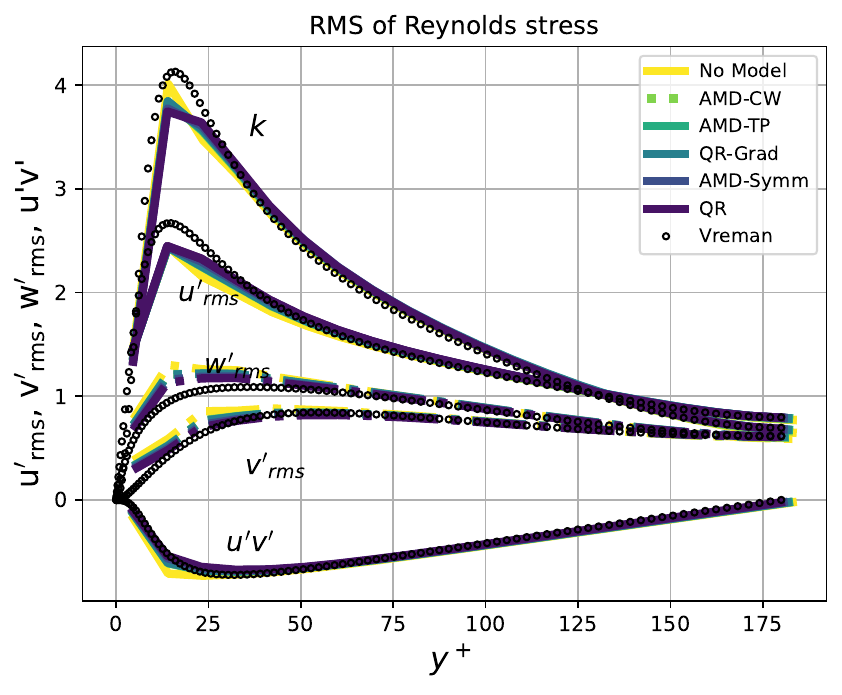} 
	\includegraphics[width=0.32\linewidth]{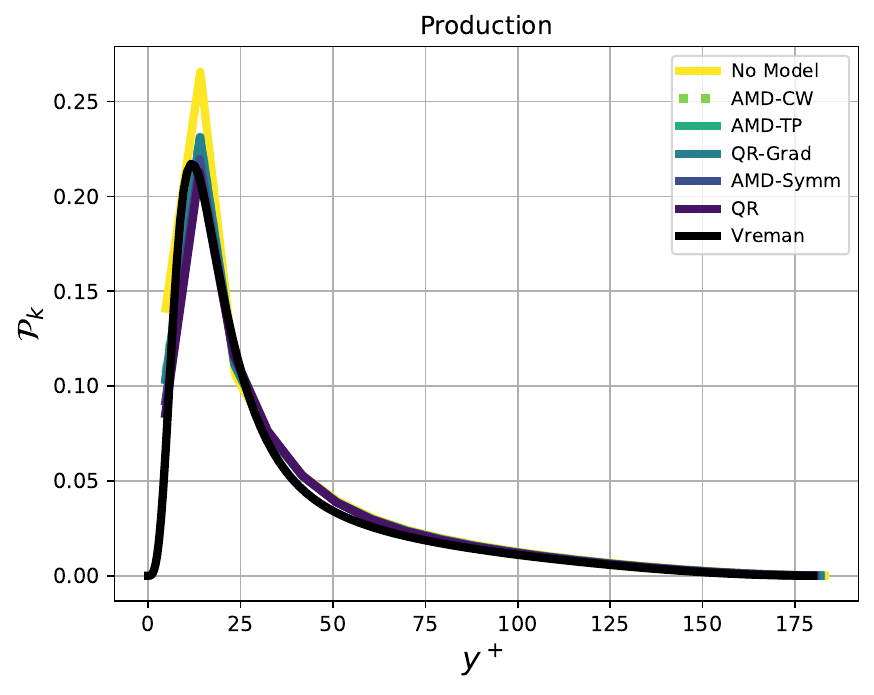}
	\includegraphics[width=0.32\linewidth]{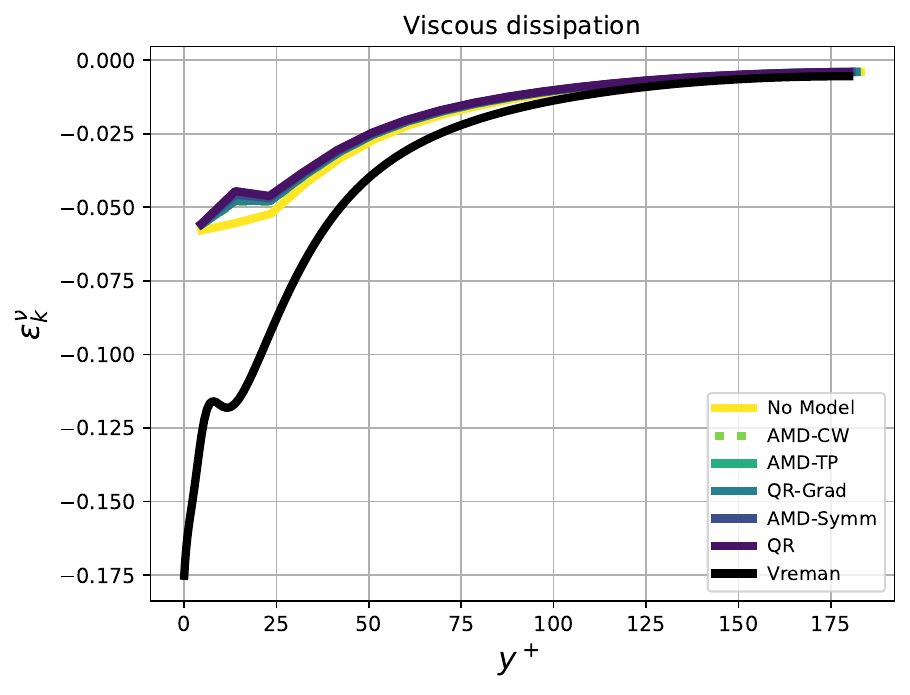} 
	\includegraphics[width=0.32\linewidth]{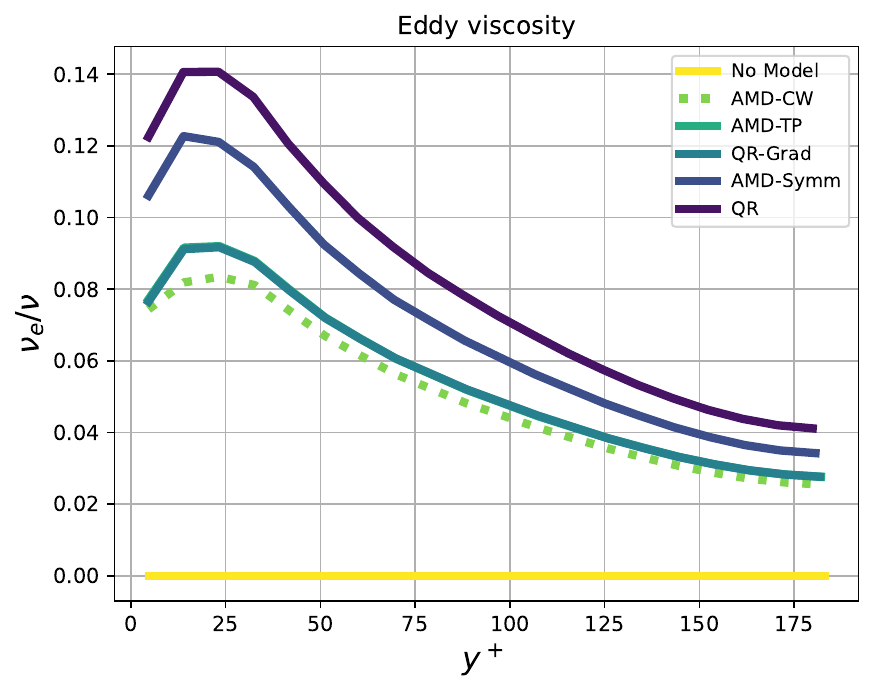} 
	\includegraphics[width=0.32\linewidth]{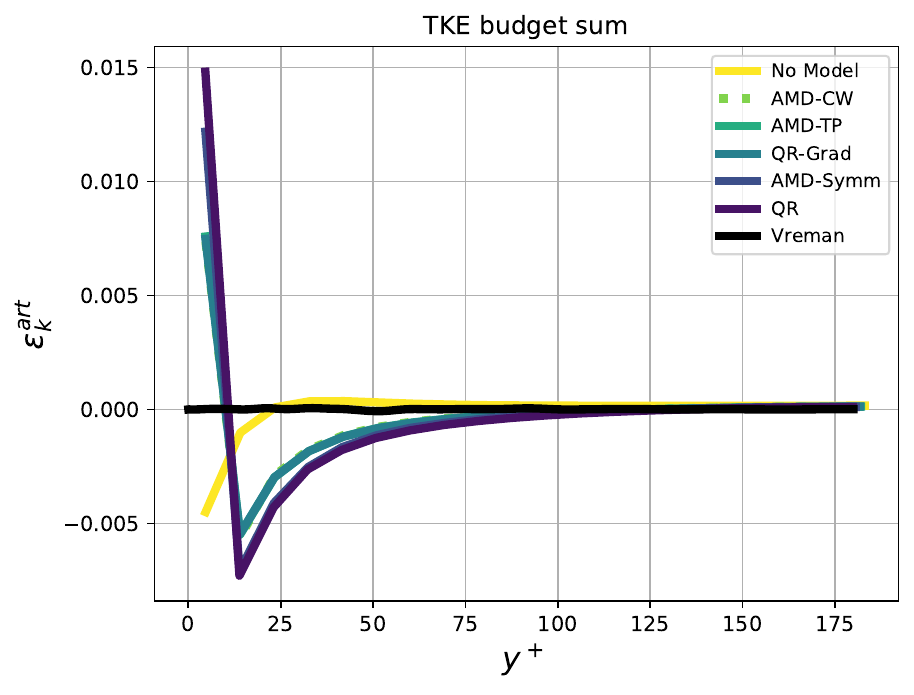} 
	\captionsetup{font = {footnotesize}}
	\caption{Channel flow simulation on isotropic grids with four minimum-dissipation models of LES and without any model. "AMD-TP" and "AMD-CW" are obtained from Eq.\eqref{eq:AMD} using different orders of summation in the tensor inner product.}
	\label{fig:iso1}
\end{figure} 
The size of the computational domain for the channel flow on isotropic mesh is $4\pi \times 2 \times 4\pi/3$ with uniform grid size $\Delta x= \Delta y = \Delta z = 0.05$.

The model coefficient is set to $C = 0.052$ to ensure a fair comparison.  Note that if the model coefficient $C$ is set to $\frac{1}{12}$, the AMD model is essentially identical to the Clark model \cite{clark1979evaluation}. The statistics are averaged over 180FTT time after the flow has developed for 20FTT time.

The profiles of the flow variables are depicted in Figure \ref{fig:iso1}. The results from simulations with SGS models consistently outperform those without an SGS model. The predictions of flow quantities show insignificant deviation. The QR and AMD-Symm models produce nearly identical results and are more accurate in predicting $\bar u, u'_{rms}$, $v'_{rms}, w'_{rms}, u'v',$ and $k$. The results from the AMD-TP, AMD-CW, and QR-Grad models are less accurate and overlap with each other. The TKE budget terms ($\mathcal P_k, \mathcal T_k, \mathcal D_k^\nu, \epsilon^\nu_k, \mathcal D_k^p$) predicted by the five versions of the minimum-dissipation models are nearly identical. 

Interestingly, the five versions of the minimum-dissipation models yield the same distribution but different amounts of eddy viscosity $\nu_e$, as shown in Figure  \ref{fig:iso1}. 
In all cases, the eddy viscosity is concentrated near the wall region, where turbulence is most vigorous and small-scale vortices abound. The extremely coarse grid ($\Delta y^+ =9$) resolves mostly large-scale motions, leaving excessive energy in the sub-grid scales, i.e., turbulent kinetic energy. Therefore, in the near-wall region, the unresolved turbulence carries too much energy, necessitating additional artificial damping to maintain local equilibrium.

The minimum-dissipation models switch off at the no-slip wall, meaning $\nu_e$ increases from zero at the wall. Figure \ref{fig:iso1} shows $\nu_e$ at the cell center; hence the $\nu_e$ profile starts from a non-zero value. It then peaks in the range $15 < y^+ < 25$ and monotonically decreases to a much lower level at the channel center, where the flow is nearly isotropic, and kinetic energy is primarily carried by large-scale motions. Regarding the amount of eddy viscosity,  the QR model introduces the largest $\nu_e$, followed by AMD-Symm with a lower amount, then AMD-TP and QR-Grad, which show identical and even lower predictions. The AMD-CW model predicts the smallest amount of $\nu_e$, about half that of the QR model.  

For the rate of artificial dissipation $\epsilon^{art}_k$, the no-model simulation significantly deviates from those with SGS models. In the region $y^+ > 13$, simulations with SGS models resemble the no-model simulation but are shifted away from the wall.  In the near-wall region $y^+ < 12$, $\epsilon^{art}_k$ from the no-model simulation is negative, indicating dissipation of turbulent kinetic energy. In contrast, $\epsilon^{art}_k$ from the simulations with SGS models is positive, indicating the production of turbulent kinetic energy. This production in the region $y^+ < 12$ stems from two mechanisms: the transfer of kinetic energy from resolved scales to unresolved scales, and the mitigation of molecular dissipation, leading to a pileup of sub-grid kinetic energy.
 
\subsection{Optimal QR Model Coefficient at Re$_\tau$ = 180} 
\label{sec:QRcoeff}
\begin{table}[t!]
	\footnotesize    
	\centering
    \begin{tabular}{cc}    \hline\hline
    \setlength\tabcolsep{3em}
    Domain  &  $4\pi \times 2 \times 4\pi/3$\\
    Mesh    & $48\times 76 \times 48$ \\
    $\Delta x^+$ and $\Delta z^+$ & 47 and 16 \\
    $\Delta y^+_w$ and $\Delta y^+_b$ &0.9 and 14\\
    $N_{y^+\leq 5}$&4\\
    Max CFL &   0.4 \\
    Developing time &   20FTT \\
	Averaging time &	180FTT (if not explicitly specified)\\
    Temporal Scheme &   Crank-Nicolson \\
    Spatial schemes  &   Symmetry-preserving\\
    Pressure correction  &  van Kan method\\     \vspace{-0.8em}\\
	SGS model&		Various models\\          \vspace{-0.8em}\\
    \multirow{2}{*}{Solver}  &   Pressure: GAMG \\  &  Velocity: PBiCG \\   \vspace{-0.8em}\\
    \multirow{2}{*}{Preconditioner}  & Pressure : DICGaussSeidel \\
                            &   Velocity: DILU \\   \vspace{-0.8em}  \\                       
    \multirow{2}{*}{PISO interation}   &Predictor step:  2\\    &Corrector step:  4 \\
    \hline\hline
    \end{tabular}
    \captionsetup{font=footnotesize}
    \caption{Numerical settings for optimizing coefficients in the QR (Sec \ref{sec:QRcoeff}) and AMD (Sec \ref{sec:AMDcoeff}) models, as well as for comparing various SGS models (Sec \ref{sec:sgsRK}) using symmetry-preserving discretization. Definitions: $\Delta y^+_w$ is the mesh size next to the wall, $\Delta y^+_b$ is the mesh size at the centerline, $N_{y^+\leq5}$ denotes the number of mesh points in the region where $y^+ \leq 5$.}
    \label{tab:numSetting}
\end{table}
\begin{figure}[b!]
	\centering 
	\includegraphics[width=0.32\linewidth]{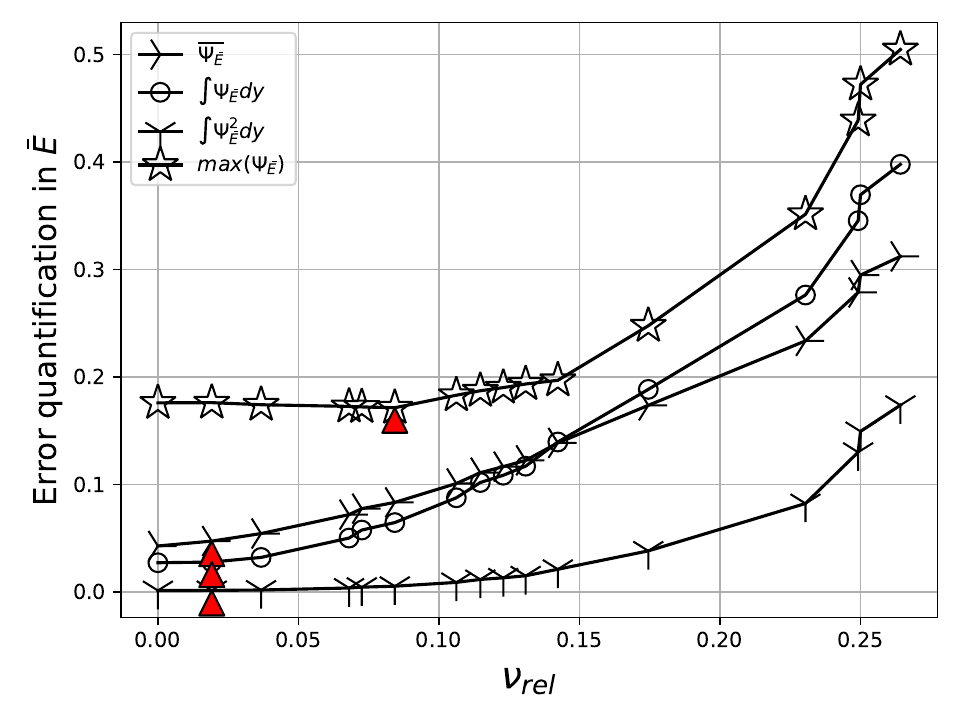}
	\includegraphics[width=0.32\linewidth]{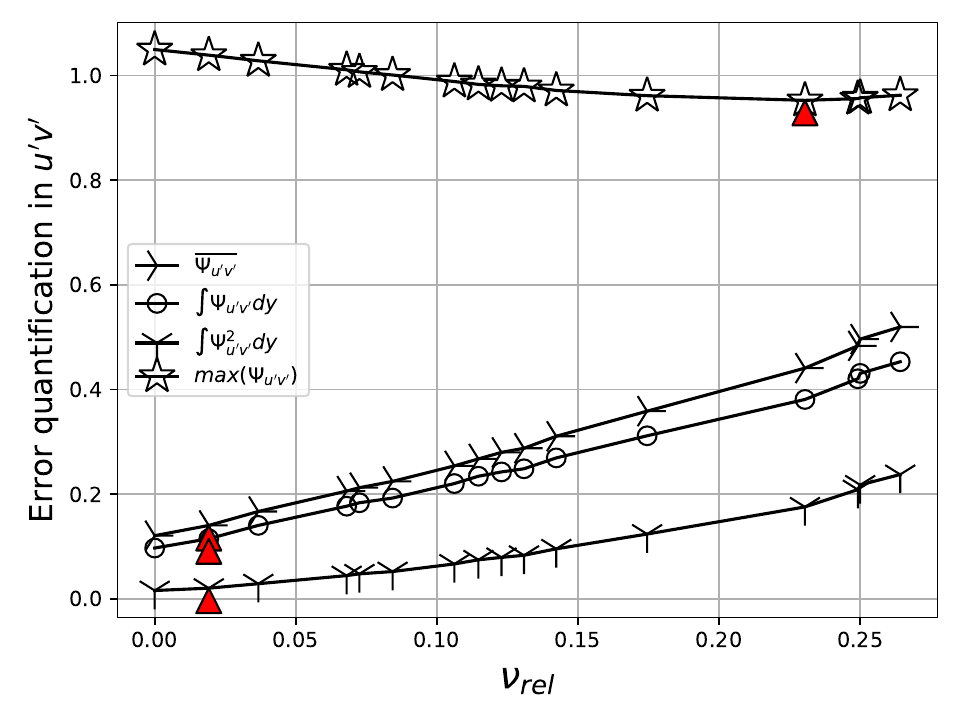}             
	\includegraphics[width=0.32\linewidth]{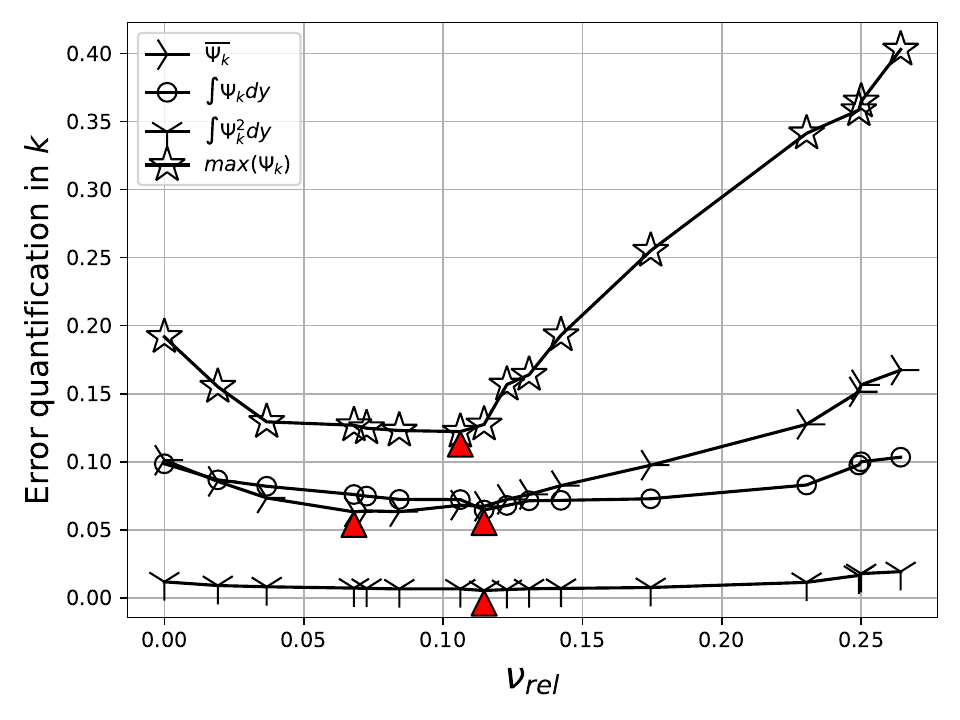} 
	\captionsetup{font = {footnotesize}}
	\caption{Error quantification using the kinetic energy of the mean flow $\bar E$, shear stress $u'v'$, and turbulent kinetic energy $k$ in a channel flow simulation at $\mathrm{Re}_\tau = 180$ while optimizing the QR coefficient.}
	\label{fig:EQQRcoeff}
\end{figure} 
The model coefficients are selected across a broad span, ranging from $C=0.012$ to $C=0.471$. Coefficients exceeding 0.062 are determined through literature references \cite{verstappen2018much,verstappen2011does,rozema2015minimum,Rozema2020LowDissipationSM}, while the remaining values are guided by consistency with the Smagorinsky model and practical insights. Detailed numerical configurations are provided in Table \ref{tab:numSetting}.

We analyze the relative differences for key flow quantities: kinetic energy of the mean flow $\bar E$, Reynolds shear stress $u'v'$, turbulent kinetic energy $k$, and enstrophy $\xi$. This assessment employs four distinct methodologies, detailed in Figure \ref{fig:EQQRcoeff}.
Each figure in the series includes four error metrics for a specific flow quantity, pinpointing the minimum error for each metric using red triangles. The error analysis yields the following insights:\\
1) Among the evaluated coefficients, $C=0.012$ provides the most accurate representation of  Reynolds shear stress $u'v'$, and the kinetic energy of the mean flow $\bar E$. Therefore, for accurately modeling Reynolds shear stress $u'v'$, and the large-scale motions, the preferred coefficient is $C= 0.012$.\\
2) QR coefficients C=0.092 demonstrate minimal errors in the turbulent kinetic energy $k$. Consequently, for capturing small-scale motions effectively, the optimal coefficient is identified as $C=  0.092$.\\
The optimal coefficient $C=0.012$ appears to be smaller than expected. Recall that the precursor (no-model) simulation is optimized so that the artificial dissipation $\epsilon^{art}_k$  is nearly zero. Adding SGS models to an artificial dissipation-free case will likely disrupt the existing global equilibrium. In practice, the simulations are not optimized as here. Therefore larger values, for instance $C\leq0.052$ ($\nu_{rel}\leq0.1$), are also capable of accurately predicting the large scale motions $\bar E$ as shown in Figure \ref{fig:EQQRcoeff}. Note that the turbulent kinetic energy is lumped into the pressure in the NS equation, meaning that $k$ is not modeled via SGS models. Thus, the error metrics for $k$ are less indicative than that of the Reynolds shear stress $u'v'$ and the kinetic energy of the mean flow $\bar E$.
\subsubsection{Turbulent Characteristics}
\begin{figure}[!b]
	\centering
	\includegraphics[width=0.32\linewidth]{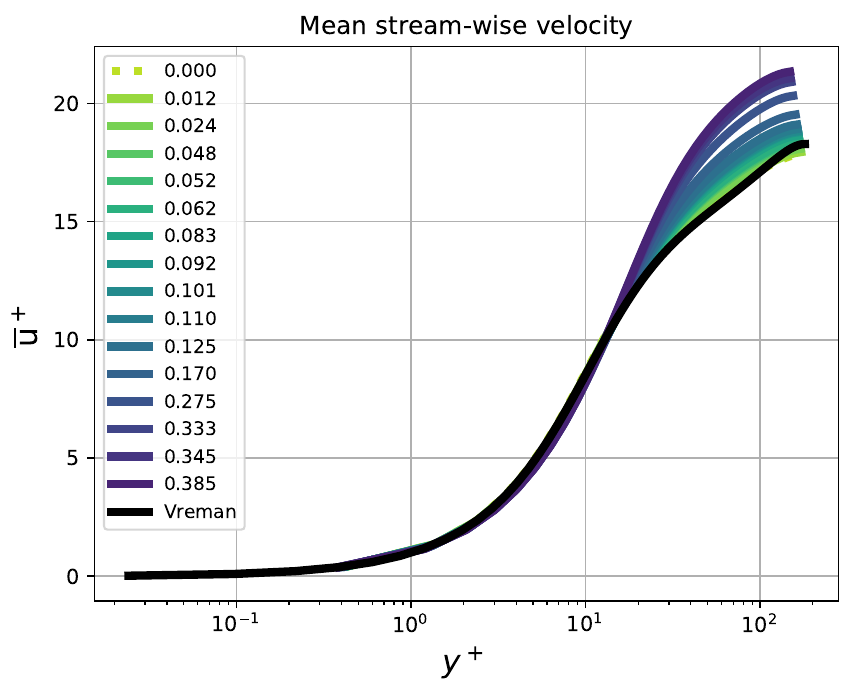}
	\includegraphics[width=0.32\linewidth]{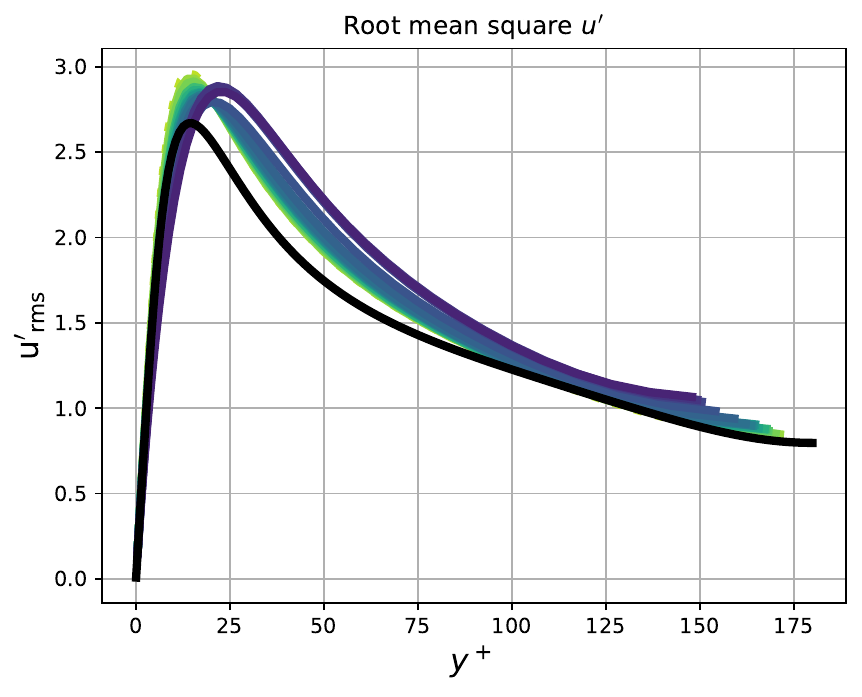} 
	\includegraphics[width=0.32\linewidth]{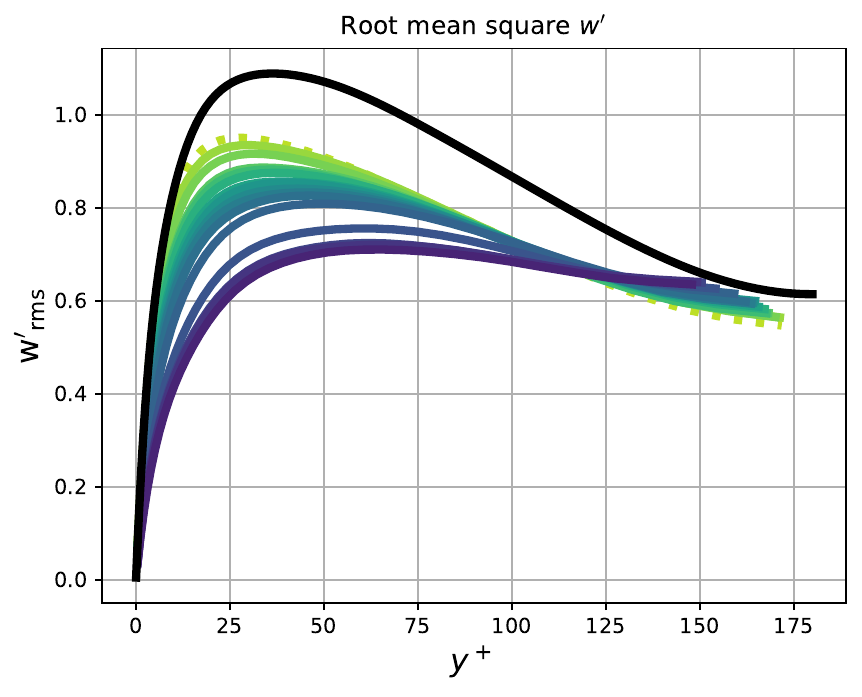} 
	\includegraphics[width=0.32\linewidth]{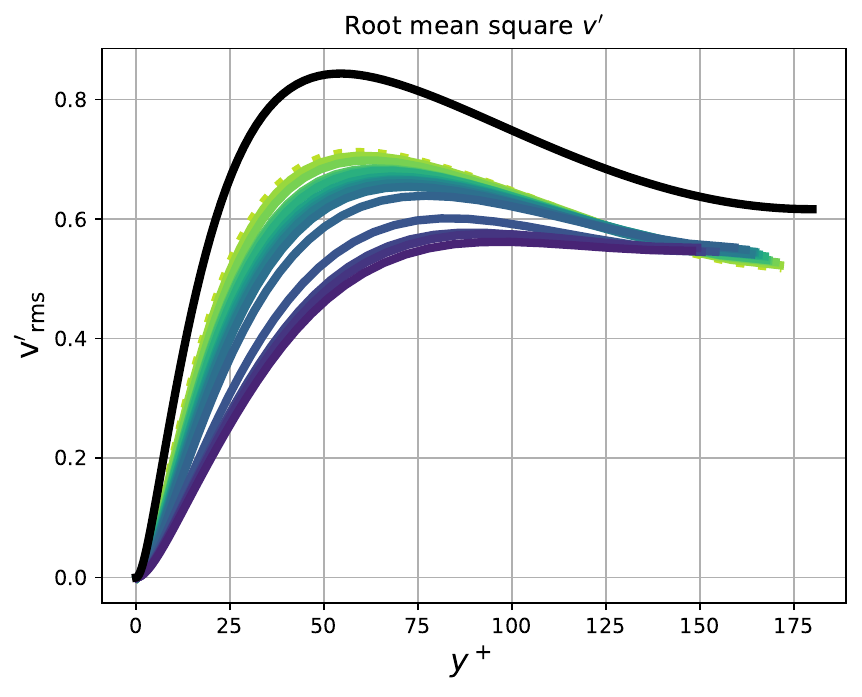} 
	\includegraphics[width=0.32\linewidth]{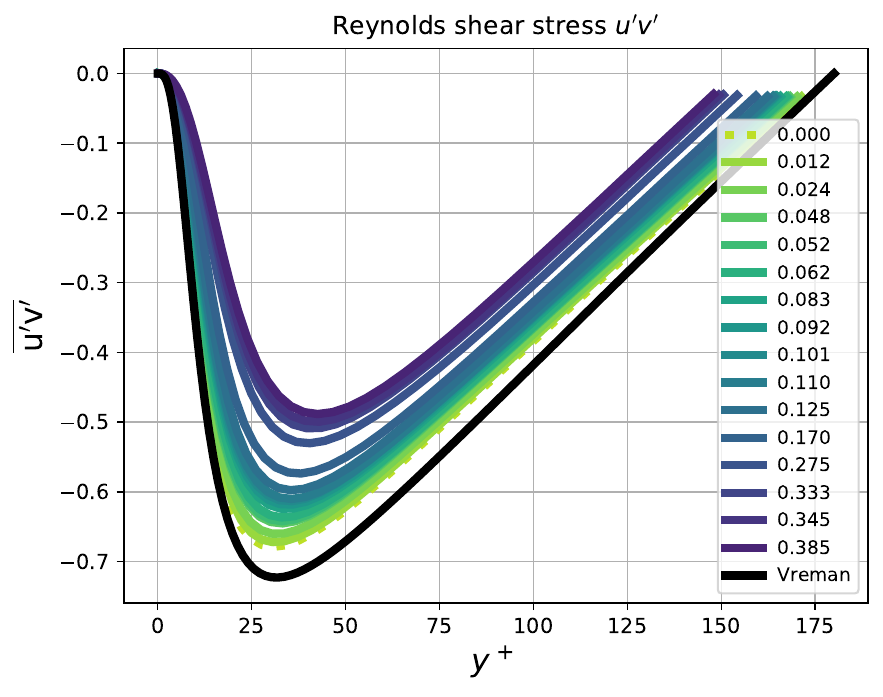} 
	\includegraphics[width=0.32\linewidth]{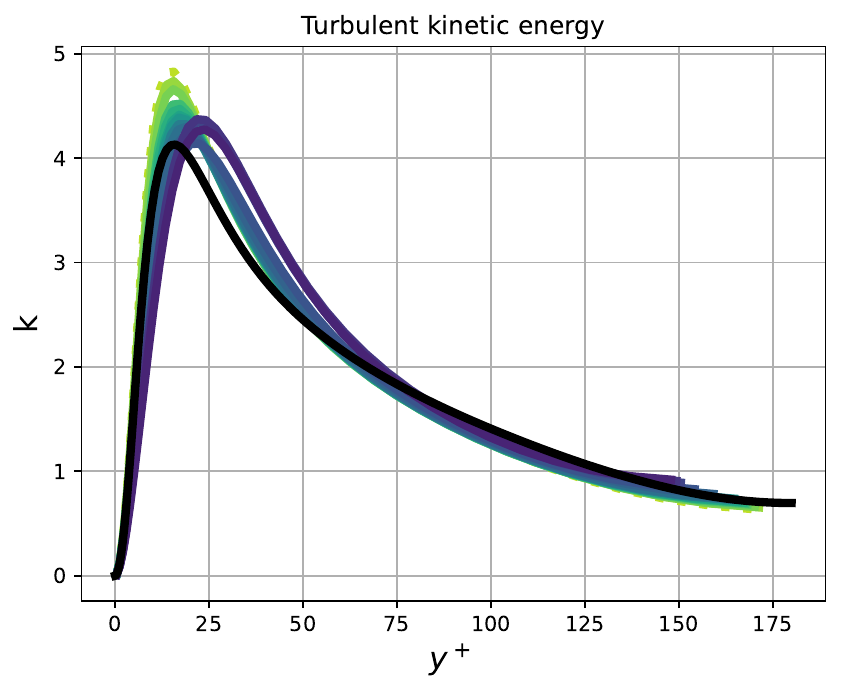} 
	\captionsetup{font = {footnotesize}}
	\caption{	Mean velocity, velocity fluctuation, and turbulent kinetic energy versus wall-normal distance in channel flow simulations at $\mathrm{Re}_\tau = 180$ for optimizing the QR model coefficient using symmetry-preserving discretization.}
	\label{fig:QRcoeff1}
\end{figure}
\begin{figure}[!b]
	\centering
	\includegraphics[width=0.32\linewidth]{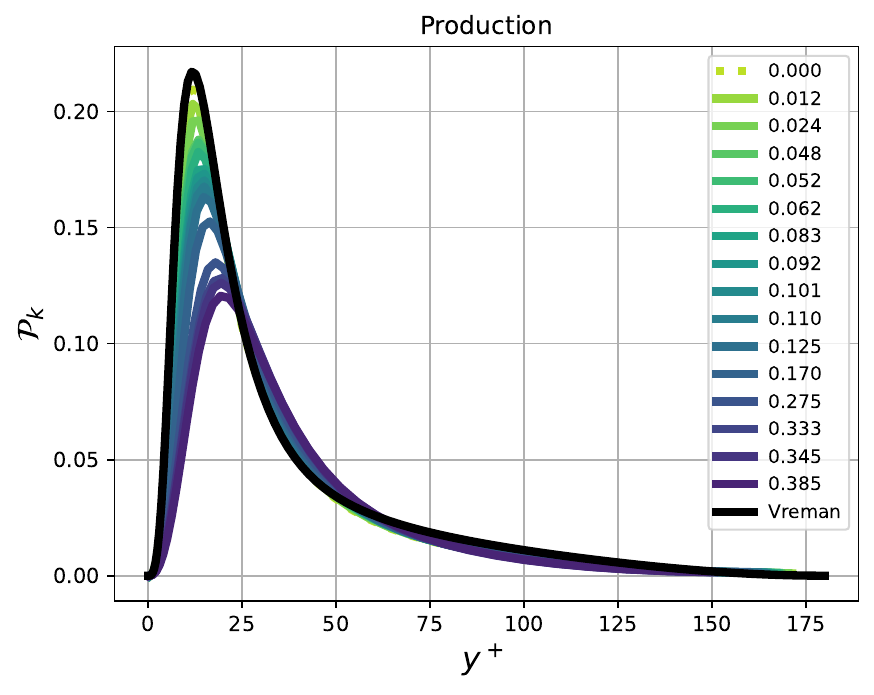}
	\includegraphics[width=0.32\linewidth]{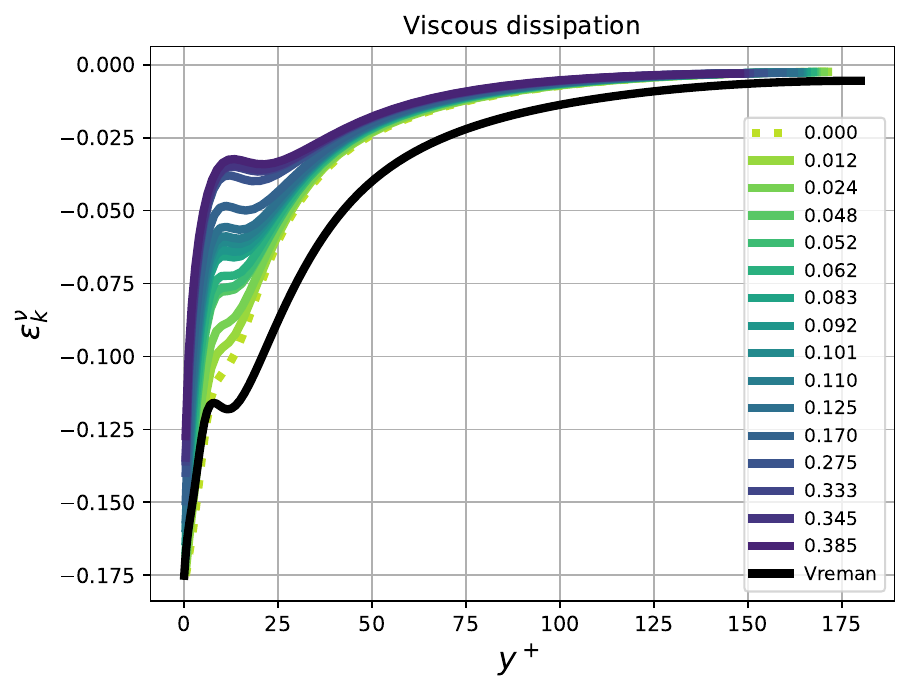}
	\includegraphics[width=0.32\linewidth]{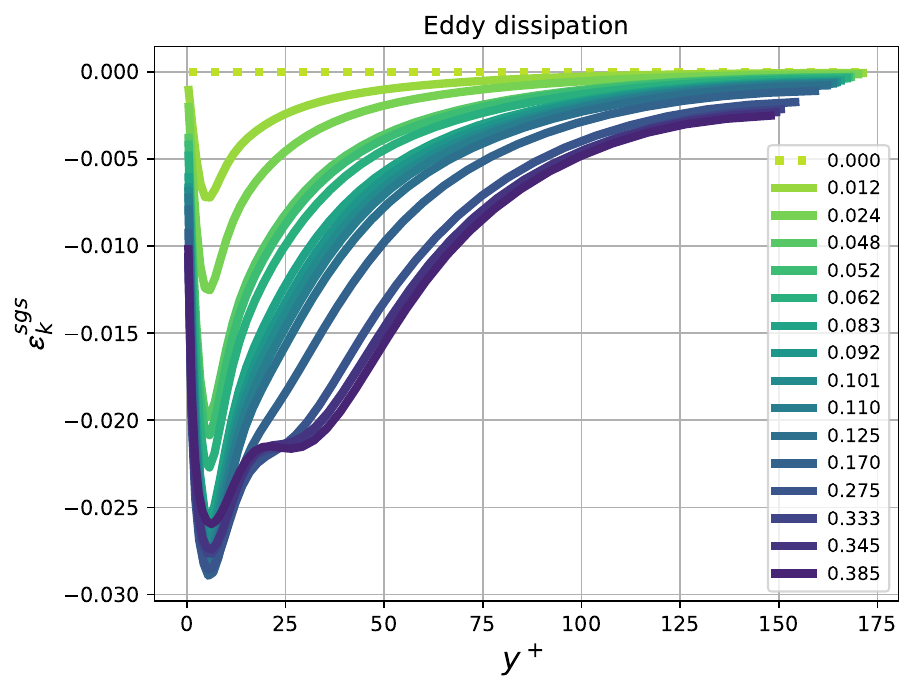}
	\includegraphics[width=0.32\linewidth]{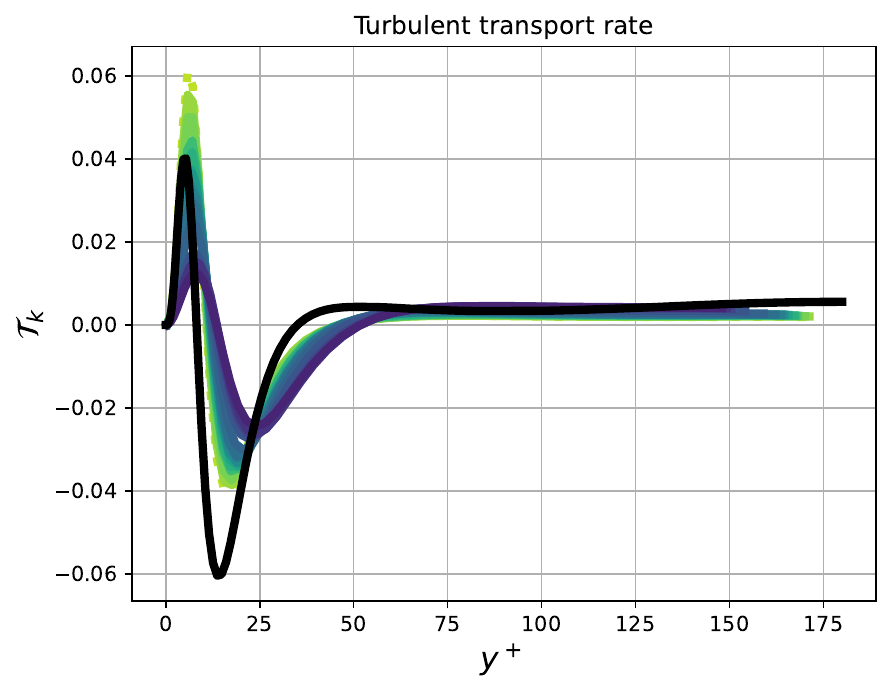}
	\includegraphics[width=0.32\linewidth]{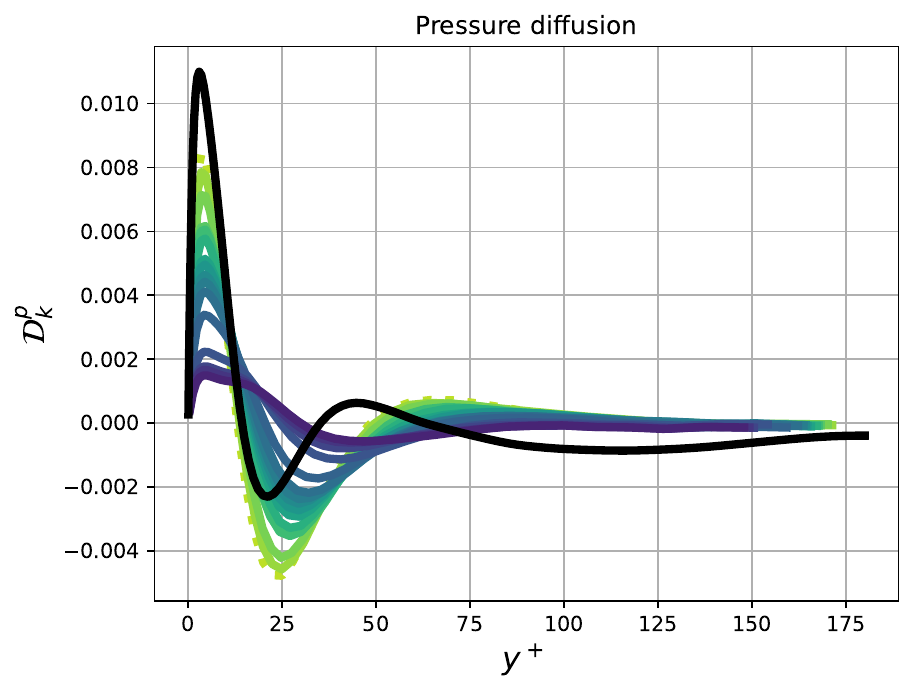}
	\includegraphics[width=0.32\linewidth]{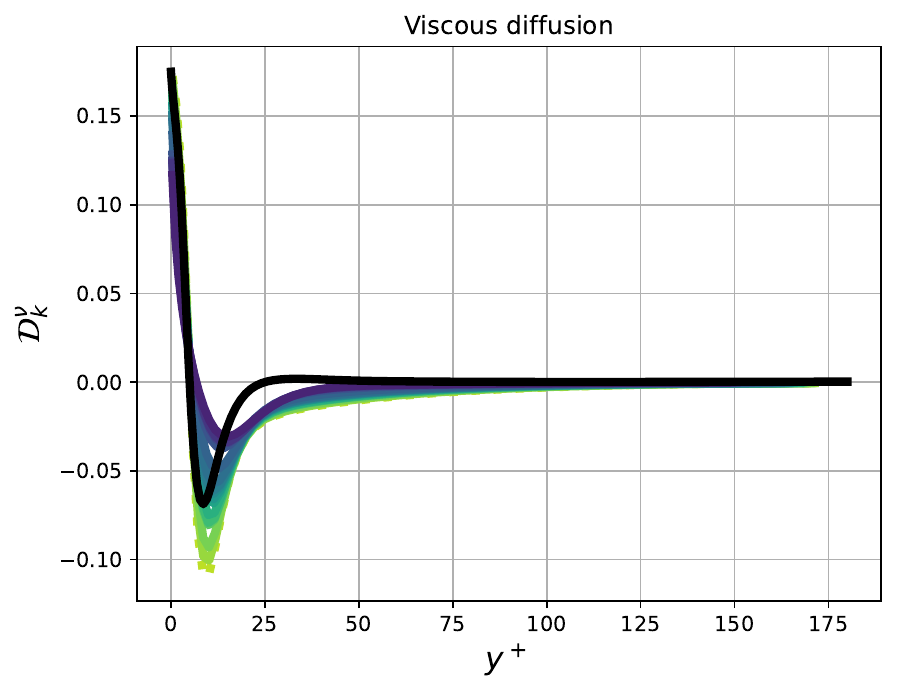}
	\includegraphics[width=0.32\linewidth]{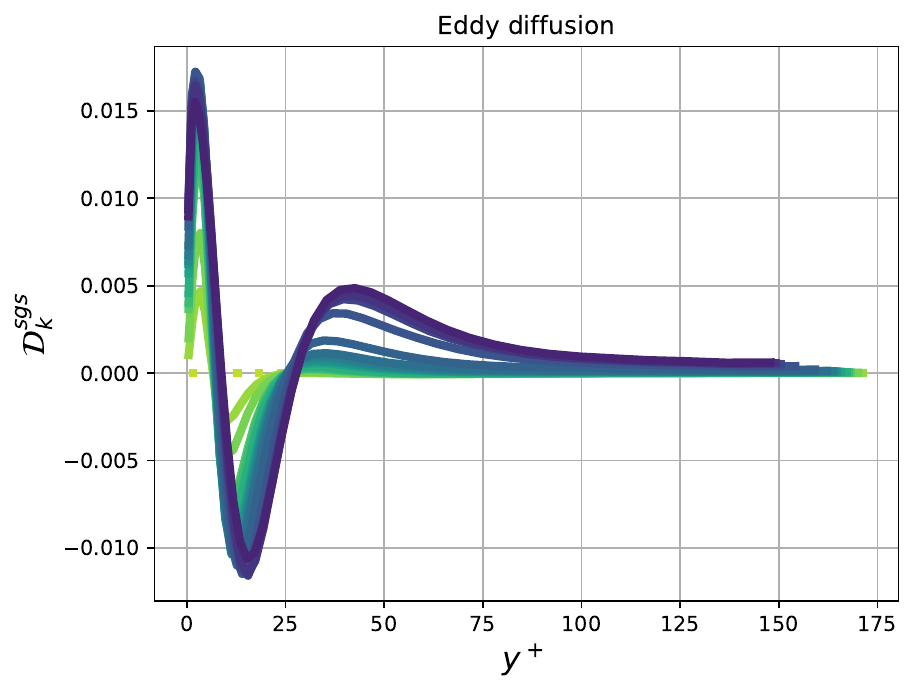}
	\includegraphics[width=0.32\linewidth]{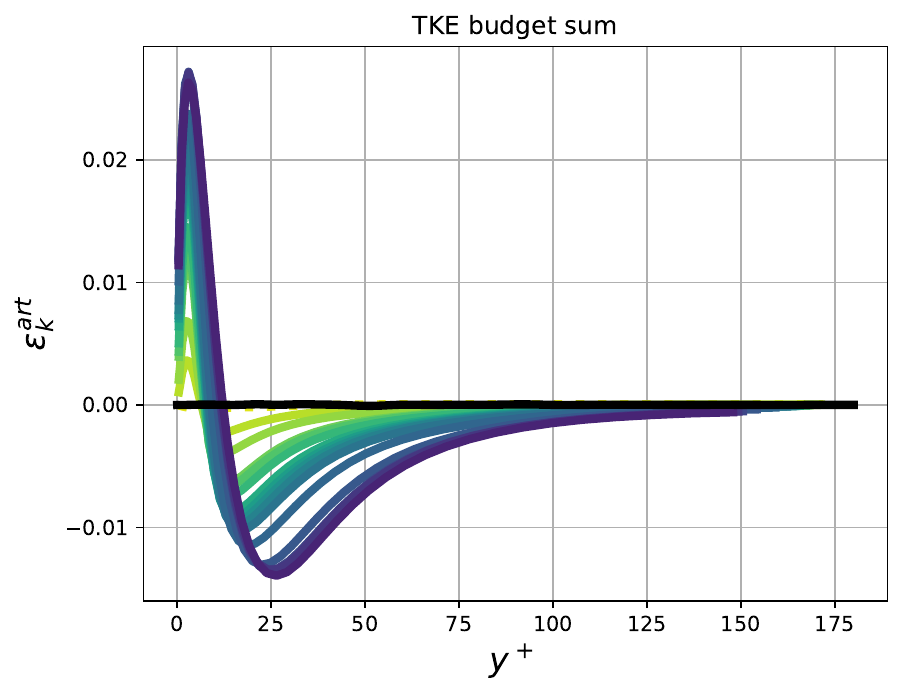}
	\includegraphics[width=0.32\linewidth]{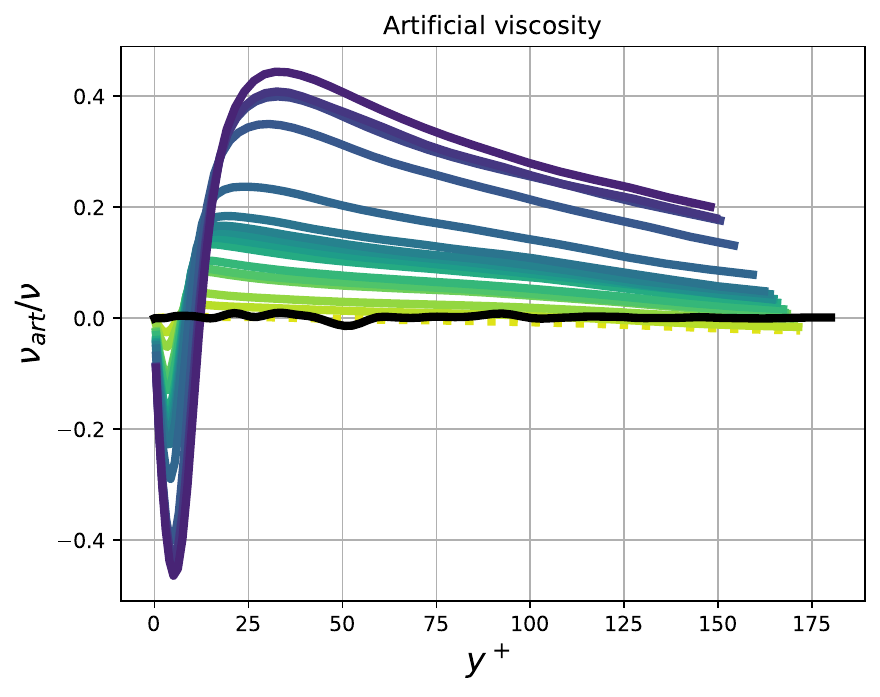}   
	\captionsetup{font = {footnotesize}}
	\caption{Energy balance terms versus wall-normal distance in channel flow simulations at $\mathrm{Re}_\tau=180$ for optimizing the QR model coefficient using symmetry-preserving discretization. }
	\label{fig:QRcoeff2}
\end{figure}
\begin{figure}[!b]
	\centering 
	\includegraphics[width=0.32\linewidth]{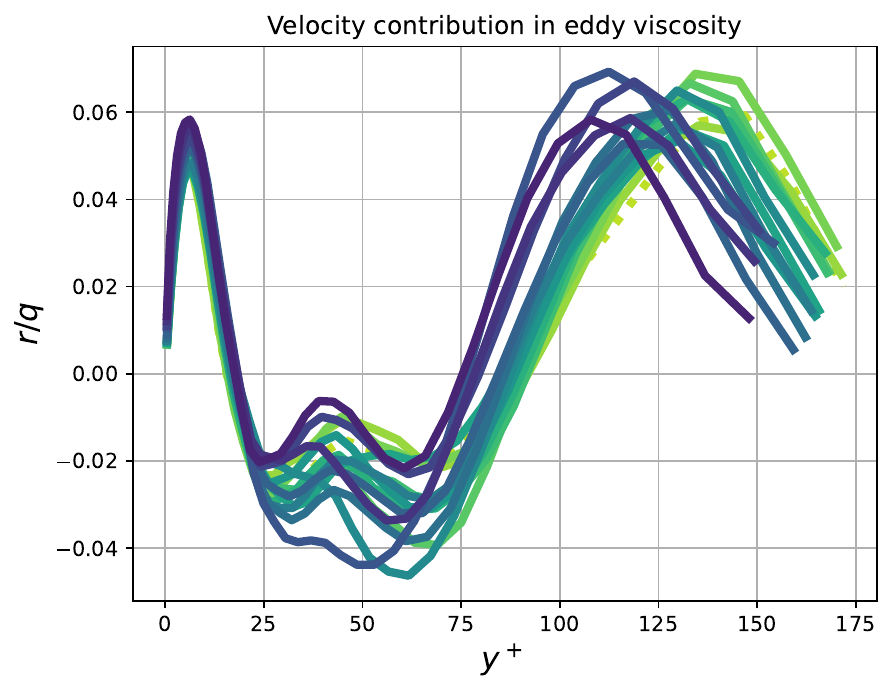} 
	\includegraphics[width=0.32\linewidth]{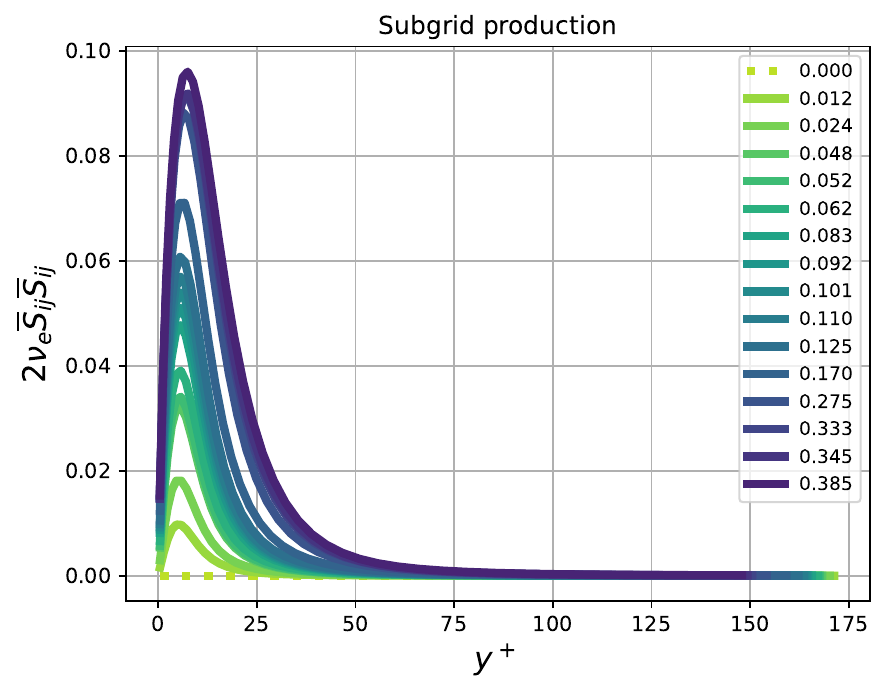}
	\includegraphics[width=0.32\linewidth]{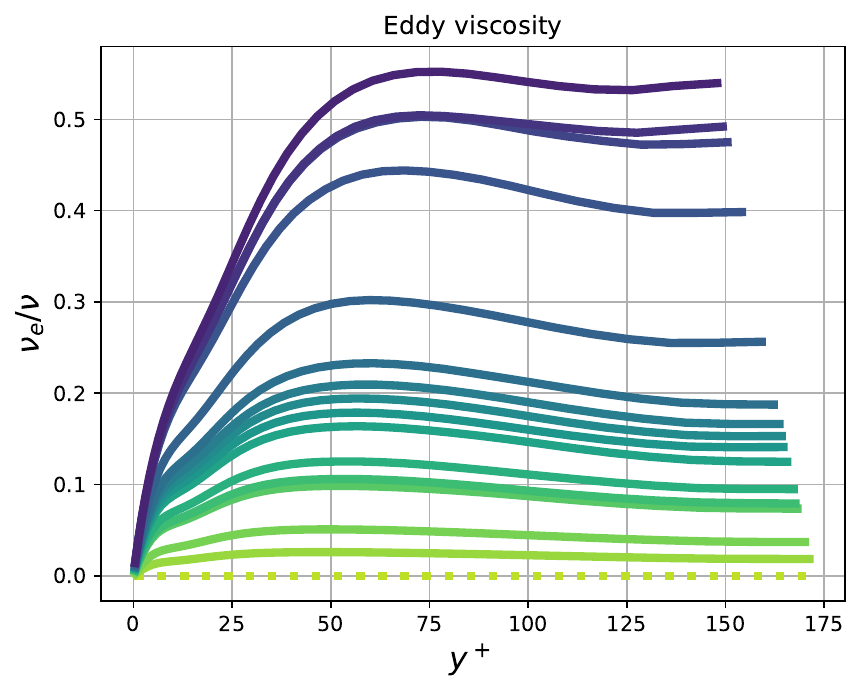}  
	\captionsetup{font = {footnotesize}}
	\caption{Subgrid-scale model quantities versus wall-normal distance in channel flow simulations at $\mathrm{Re}_\tau=180$ for optimizing the QR model coefficient using symmetry-preserving discretization. }
	\label{fig:QRcoeff5}
\end{figure}
The profiles of streamwise mean velocity $\bar u$, RMS of $u'u'$, $v'v'$ and $w'w'$, shear stress $u'v'$, turbulent kinetic energy $k$ are depicted against wall-normal distance in Figure \ref{fig:QRcoeff1}. The terms in the energy balance equation and the subgrid activity are shown in Figure \ref{fig:QRcoeff2} and \ref{fig:QRcoeff5}, respectively.  From these, several conclusions can be drawn:\\
1) Mean streamwise velocity $\bar u$: 
In the near-wall region, predictions with different coefficients closely match the DNS data, indicating that all considered coefficients accurately capture near-wall flow behaviors, crucial for yielding correct predictions in the outer layer of the wall. In the outer wall layer $0.2\delta < y < \delta$, increasing the coefficient from $C = 0.012$ to 0.385 results in progressively increased over-predictions of the mean streamwise velocity. As the QR coefficient increases, the mean streamwise velocity $\bar u$ exhibits less flatness, showing a tendency toward laminar flow. \\
2) RMS of Reynolds stress components and turbulent kinetic energy:
The RMS of streamwise $u'u'$, wall-normal $v'v'$, and spanwise $w'w'$ components of Reynolds stress, along with the turbulent kinetic energy $k$, demonstrate a progressive shift of peak values away from the wall with increasing QR coefficient. Significant underpredictions are observed in the peak values of  $w'w'$, $v'v'$, and $u'v'$, while equivalent overpredictions are evident in the peak values of $u'u'$ and $k$, suggesting reduced turbulent energy transfer from the streamwise to the spanwise and wall-normal components.  Changes in $v'v'$ and $w'w'$ occur more slowly across the computational domain, whereas changes in $u'u'$ and $k$ increase more slowly in the near-wall region ($y^+<25$) but decrease more rapidly in the region $y^+>75$. \\
3) Turbulent kinetic energy balance terms:
As the QR coefficient $C$ increases, the production $\mathcal P_k$, molecular dissipation $\epsilon^{\nu}_k$, turbulent transport $\mathcal T_k$, pressure diffusion $\mathcal D_k^p$ and viscous diffusion $\mathcal D_k^v$ rates progressively decrease, while the eddy-related terms $\epsilon^{sgs}_k$ and $\mathcal D_k^{sgs}$ progressively increase in the region $y^+<70$. Beyond $y^+>75$, the budget terms, except for $\epsilon^{sgs}_k$, exhibit self-similarity. That is,  the profiles of each budget term plotted against the wall-normal distance collapse for all $y^+$ beyond the development region.
The $\mathcal P_k$, $\epsilon^{\nu}_k$ and the eddy dissipation $\epsilon^{sgs}_k$ account for the net increase and decrease of the subgrid kinetic energy. The peaks of $\mathcal{P}_k$ and $\epsilon^{\nu}_k$ decrease to  $70\% $ and $30\% $, respectively, of their reference DNS values. 
The redistribution terms $\mathcal{T}_{k}$, $\mathcal D_k^p$, and $\mathcal D_k^v$ become smoother and approach zero, whereas $\mathcal D_k^{sgs}$ become more and more intense dominating the redistribution process.
The rate of artificial dissipation $\epsilon^{art}_k$ exhibits two opposite behaviors, producing sub-grid kinetic energy in the near wall region $y^+<12$ and dissipating the turbulent kinetic energy in the region $20<y^+<125$. \\
4) Eddy viscosity $\nu_e$: The eddy viscosity gradually ascends from zero at the no-slip wall to the maximum at $y^+\approx 50$. This distribution sharply contrasts with the results from isotropic computations shown in Figure \ref{fig:iso1}, where the eddy viscosity is predominantly introduced in the near-wall region ($y^+ \leq 40$). The maximum eddy viscosity obtained with the largest QR coefficient approximately remains half of the molecular viscosity. Error quantification shown in Figure \ref{fig:EQQRcoeff} indicates that the optimal eddy viscosity should be $5\%$ to $15\%$ of the molecular viscosity (given the current spatial and temporal resolution), specifically $0.012 \leq C \leq 0.093$. \\
5) artificial viscosity $\nu_{art}$:
The artificial viscosity is comparable to the eddy viscosity, but their distributions are essentially opposite. As the model coefficient increases, the magnitude of $\nu_{art}$ progressively increases to half of the molecular viscosity, exhibiting both positive and negative values. In the near-wall region ($y^+<12$), the artificial viscosity is negative, contributing to the production of turbulent kinetic energy. The sign changes abruptly at $y^+ \approx 12$, where both negative and positive values successively reach their peaks. Further away from the wall, $\nu_{art}$ increases monotonically, changing sign again only in cases with $C<0.012$. The negative artificial viscosity may mimic the backward energy cascade, which is physically realistic and required in many applications, but is typically clipped off in practice to stabilize the numerical simulation. 

\subsubsection{Local and Global Balance of Turbulent Kinetic Energy Budgets}
\begin{figure}[!b]
	\centering 
	\includegraphics[width=0.32\linewidth]{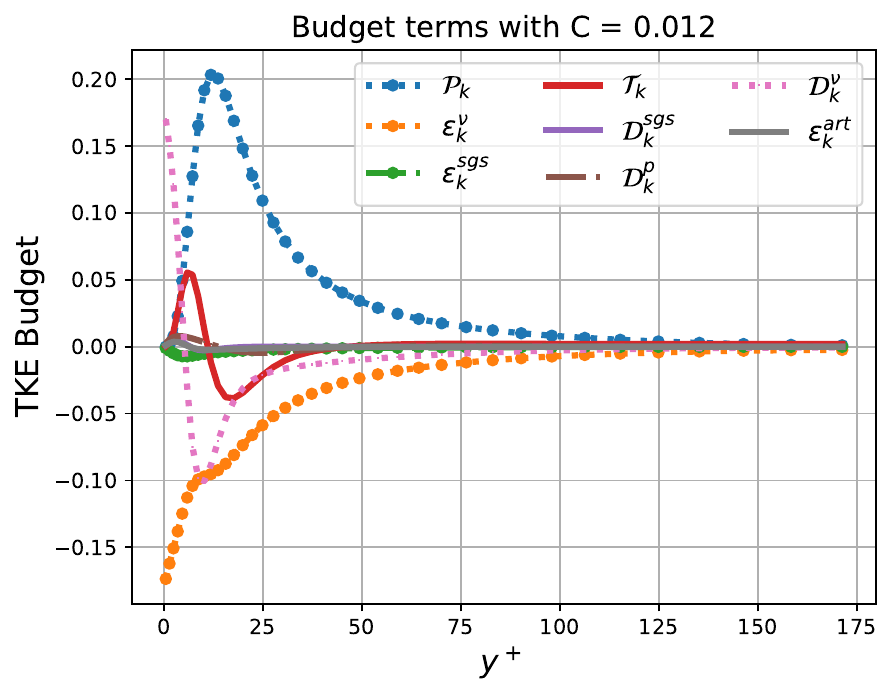} 
	\includegraphics[width=0.32\linewidth]{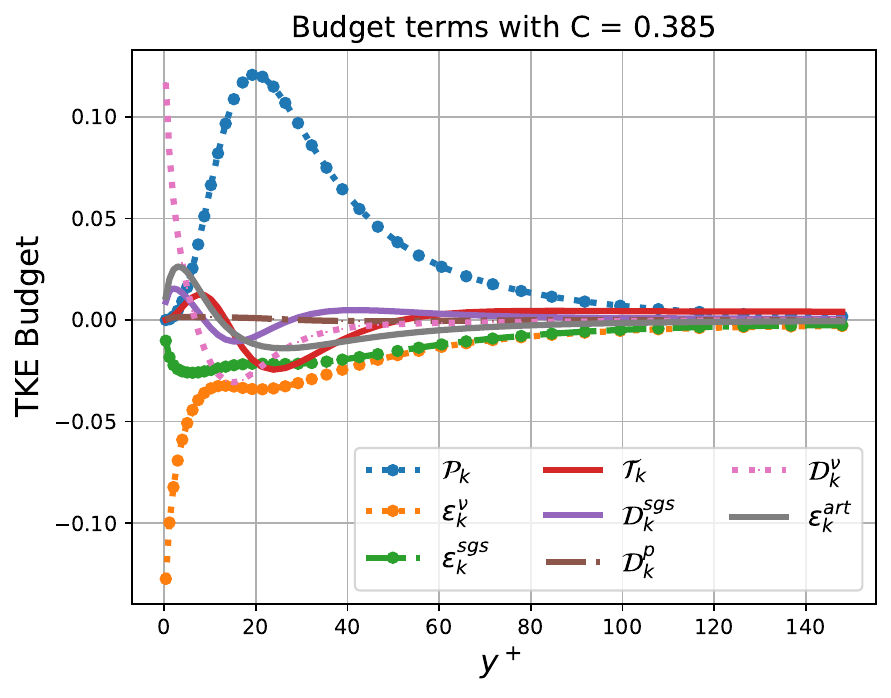} 
	\includegraphics[width=0.32\linewidth]{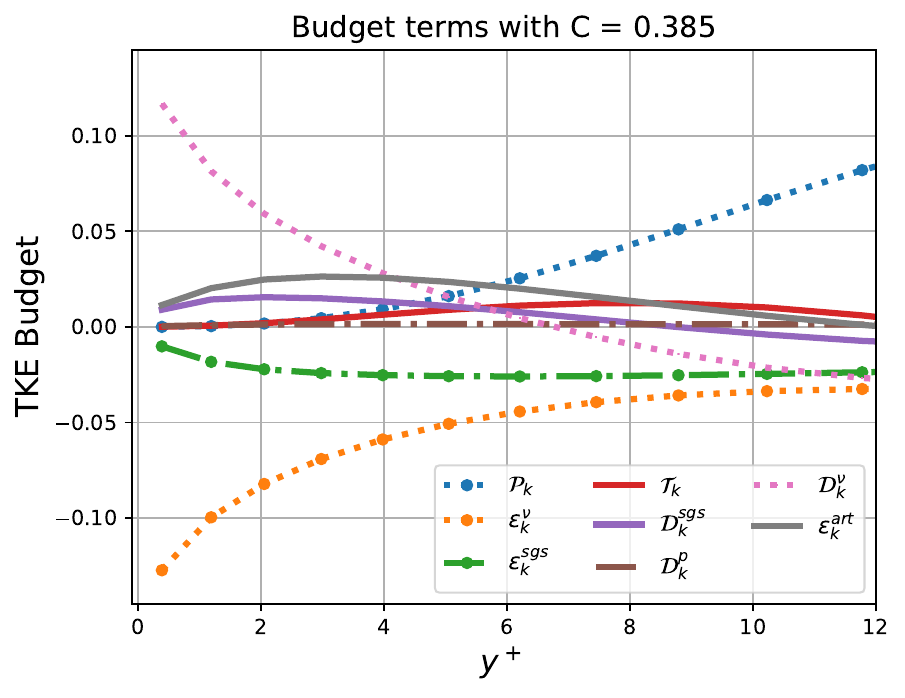} 
	\captionsetup{font = {footnotesize}}
	\caption{The budgets of turbulent kinetic energy in channel flow at $\mathrm{Re}_\tau =180$ using the QR model with $C=0.012$ and $C=0.385$ employing symmetry-preserving discretization. The budget terms are normalized by $u^4_\tau/\nu$.}
	\label{fig:QRcoeff4}
\end{figure}
\begin{figure}[!t]
	\centering 
	\includegraphics[width=0.42\linewidth]{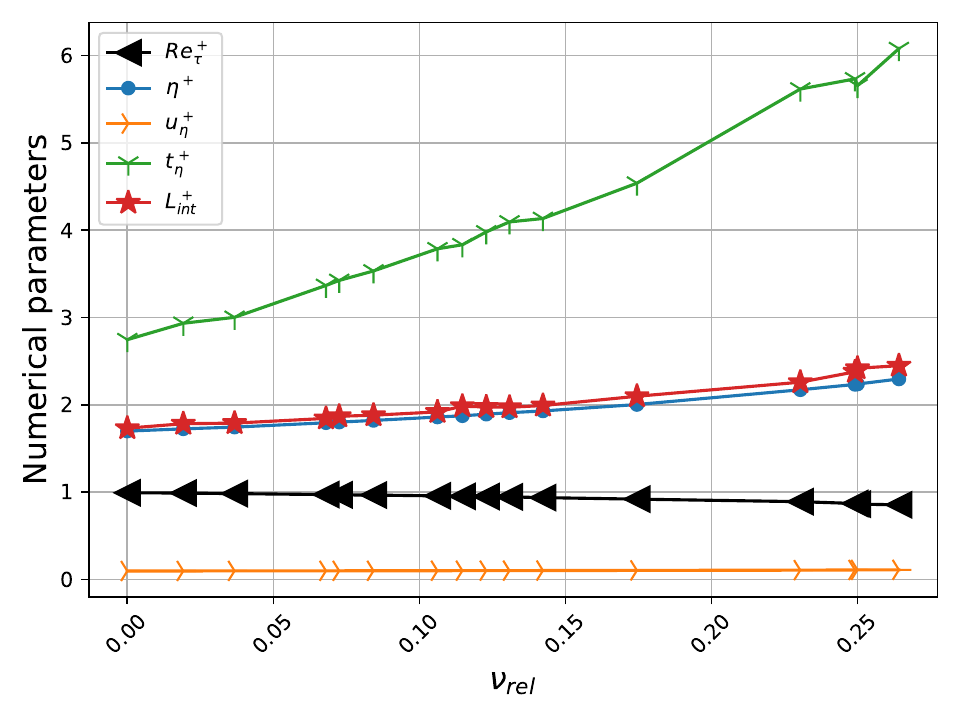} 
	\captionsetup{font = {footnotesize}}
	\caption{Numerical parameters versus relative eddy viscosity $\nu_{rel}=\frac{\langle\nu_e\rangle}{\langle\nu_e\rangle+\langle\nu\rangle}$ in channel flow simulations optimizing the QR model coefficient using symmetry-preserving discretization.}
	\label{fig:QRcoeff6}
\end{figure} 
The progressively enlarged positive value of artificial dissipation rate in the near wall region $y^+<12$ with the increase of model coefficient shown in Figure \ref{fig:QRcoeff2} can be explained by examining the local and global balance of TKE budgets.

Figure \ref{fig:QRcoeff4} shows the local energy balance versus wall-normal distance, obtained from simulations with the minimum ($C=0.012$) and maximum (C=0.385) coefficients, to illustrate the impact of eddy viscosity on the local energy balance.\\
Consider C=0.385 as an illustrative case.  The right panel in Figure \ref{fig:QRcoeff4} illustrates the local behavior of the budgets within $y^+<12$, where a positive artificial dissipation rate is observed. In this region, $\epsilon^{sgs}_k$ and $\epsilon^{\nu}_k$ are negative, while $\mathcal P_k$ and $\mathcal T_k$ are positive, $\mathcal D^{sgs}_k$ and $\mathcal D^{\nu}_k$ are positive first and decrease to negative, and $\mathcal D^p_k$ is approximately zero. Therefore, the positive artificial dissipation rate in $y^+<12$ can results from the over-intensive eddy and viscous transport mechanisms and insufficient dissipation $ -(\nu+\nu_e) \overline{\partial_j u_i' \partial_j u_i'}$. The latter indicates that the gradient of the residual field $u_i'$ is too smooth for $y^+<12$.
Following the local equilibrium hypothesis underlying the eddy viscosity model $\tau(v) = -2\nu_e\bar S_{ij}$, a larger model coefficient leads to a reduction in the strain rate tensor $\bar S_{ij}$, resulting in a smoother mean velocity profile $\bar u$ due to a less steep velocity gradient field.  Additionally, with the eddy viscosity reaching $40\%$ of the molecular viscosity, turbulence suppression occurs. Consequently, there is less vigorous redistribution by the turbulent transport, pressure diffusion and viscous diffusion terms, with eddy diffusion dominating redistribution in the region $y^+<100$. This results in, as depicted in Figure \ref{fig:QRcoeff6}, a reduction in Reynolds number, a thicker boundary layer, a rise in integral length scale indicating larger eddies in the flow, and the presence of larger Kolmogorov length and time scales implying the smallest motions become larger. 
 \begin{figure}[bh!] 
	\centering 
	\includegraphics[width=0.495\linewidth]{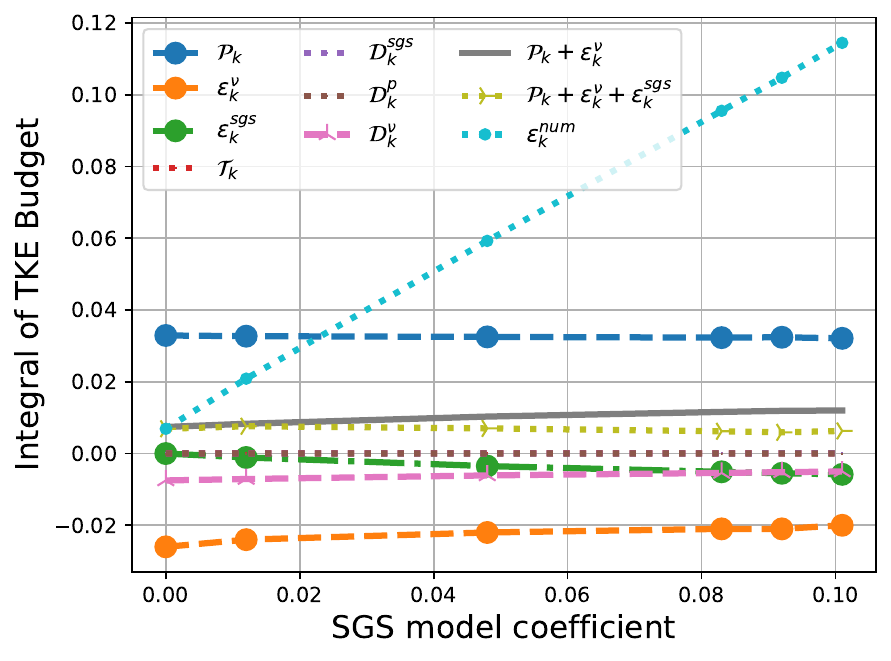} 
	\includegraphics[width=0.495\linewidth]{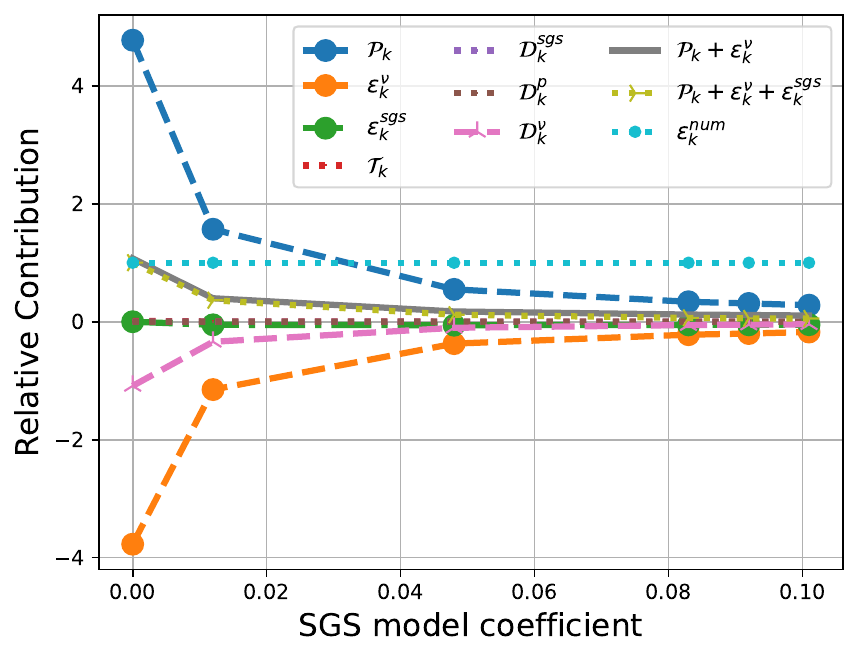} 
	\captionsetup{font = {footnotesize}}
	\caption{Integral (left) and relative contribution (right) of the TKE budget over half the channel height in a fully developed channel at $\mathrm{Re_\tau} = 180$, using different QR model coefficients with symmetry-preserving discretization.}
	\label{fig:TKEint}
 \end{figure}
The integrals and relative contributions of each TKE budget term to the artificial dissipation are presented in Figure \ref{fig:TKEint}. The relative contribution is defined as \[\frac{\int^h_0{\,\mathcal A_k}\,dy}{\int^h_0{\,\epsilon^{art}_k}\,dy}\,,\] where $\mathcal{A}_k$ represents the TKE budget terms, namely $\mathcal P_k,\epsilon^{\nu}_k,\epsilon^{sgs}_k,\mathcal T_k,\mathcal D^\nu_k,\mathcal D^{sgs}_k$, and $\mathcal D^p_k$. In the reference DNS simulation, the total integrated turbulent transport, pressure diffusion, viscous diffusion, and subgrid diffusion rates of turbulent kinetic energy ($k$) for the channel flow domain are equal to zero. Additionally, the total integrated production rate of $k$ exactly balances the total integrated dissipation rate of $k$. However,  for the no-model and QR simulations, the balance of the TKE budget can not hold due to artificial errors. 
The left panel of Figure \ref{fig:TKEint} shows that as the SGS model coefficient increases, the eddy dissipation grows, while the production and molecular dissipation decrease. The sum of the total integrated production and dissipation terms is non-zero, indicating the presence of artificial dissipation. 
The sum of the total integrated production and dissipation terms is non-zero indicating the production and dissipation rates do not balance due to artificial dissipation. It is found that the total integrated viscous diffusion $\mathcal D^\nu_k$ is negative and non-zero, counteracts the production $\mathcal P_k$ of turbulent kinetic energy. Furthermore, the artificial dissipation increases significantly with the increase of the SGS model coefficient.

The right panel of Figure \ref{fig:TKEint} illustrates the relative contributions of each TKE budget term to the artificial dissipation.  It demonstrates that production, molecular dissipation, and molecular diffusion contribute significantly to artificial dissipation. The production and dissipation terms show poor balance, resulting in a net increase in artificial dissipation. In contrast, the transport, pressure diffusion, SGS dissipation, and SGS diffusion contribute almost negligibly to the artificial dissipation and remain nearly unchanged with the increase of the SGS model coefficient. As the SGS model coefficient increases, the relative contribution of all TKE terms decreases monotonically due to the rapid rise in artificial dissipation.

\subsection{Analyzing Observed Trends in Optimizing QR Coefficients}
We have demonstrated that an increase of the QR model coefficient from $C=0.012$ to 0.385 results in progressively larger over-predictions of the mean streamwise velocity in the outer layer of the wall. 
Excessively large model coefficients yield a smoother resolved velocity gradient field, resulting in an underestimation of near-wall resolved velocity gradients and consequently a smaller friction velocity $u_\tau$. The turbulent hypothesis assume eddy viscosity has the same property as the molecular viscosity, yielding an effective viscosity $\nu_{eff} = \nu + \nu_e$. 
The reduced $u_\tau$ and larger effective viscosity jointly lead to a decrease in friction Reynolds number $\mathrm{Re}_\tau$, as depicted in Figure \ref{fig:QRcoeff6}. This decrease contributes to the observed over-predictions of mean velocity profiles away from the wall. 

In contrast to the progressively over-prediction of the outer layer of the wall, the flow behaviors in the near-wall region are accurately captured at the current mesh resolution, which is crucial for correctly predicting outer layer behaviors. 
The studies \cite{jimenez2004turbulent,flores2006effect,flores2007vorticity} demonstrate that in high-Reynolds-number turbulence the flow statistics and structure within the outer layer ($0.2\delta < y < \delta$) are relatively independently of the particular configuration of the eddies closest to the wall, even if they are partially or completely under-resolved. Drawing from this empirical conclusion, we can infer that changes in the QR model primarily affect the outer layer of the wall, indicating that the large-scale motions (energy-containing scales) are most influenced by the QR model. This finding controverts the work in \cite{LUCOR2007}, where researcher found that small scales are mainly affected by changes in the subgrid model parametric uncertainty, whereas the large-scales remain unaffected.

From an energy balance perspective, the implicit LES filter provided by a given spatial and temporal resolution necessitates a certain amount of kinetic energy damping. As our results indicate, within a consistent numerical discretization framework, the level of artificial dissipation can vary accordingly with the eddy viscosity. Consequently, a specific total non-molecular viscosity $\nu_{nm} = \nu_e + \nu_{art}$ is introduced. 

As the model coefficient $C$ increases from 0.012 to 0.275, the eddy dissipation in the near-wall region rises to a level comparable to that of molecular dissipation. Consequently, a greater proportion of turbulent kinetic energy is dissipated through eddy dissipation. Furthermore, an increase in the model coefficient dramatically reduces molecular dissipation. Changes in the Kolmogorov length-scale affirm that the smallest turbulent structures expand with increasing model coefficient. Concurrently, artificial dissipation gradually approaches half of the molecular dissipation, playing a pivotal role in damping turbulent kinetic energy. The negative artificial viscosity in the region $y^+<12$ arises from two distinct mechanisms. First, kinetic energy is transferred from the resolved scales to the unresolved scales. Second, the reduction in molecular dissipation leads to an accumulation of sub-grid kinetic energy. This negative artificial viscosity may mimic the backward energy cascade, which is physically realistic and required in many applications, but is typically clipped off in practice to stabilize the numerical simulation.

\subsection{Comparison of QR Model with Other SGS Models at Re$_\tau$ = 180} \label{sec:sgsRK}
\begin{figure}[t!]
	\centering
	\includegraphics[width=0.32\linewidth]{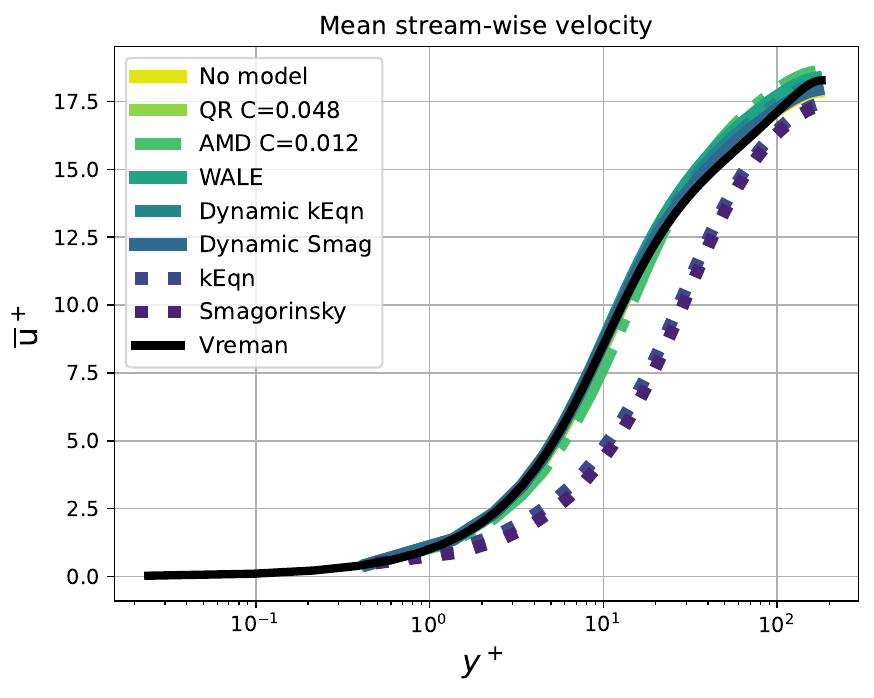}
	\includegraphics[width=0.32\linewidth]{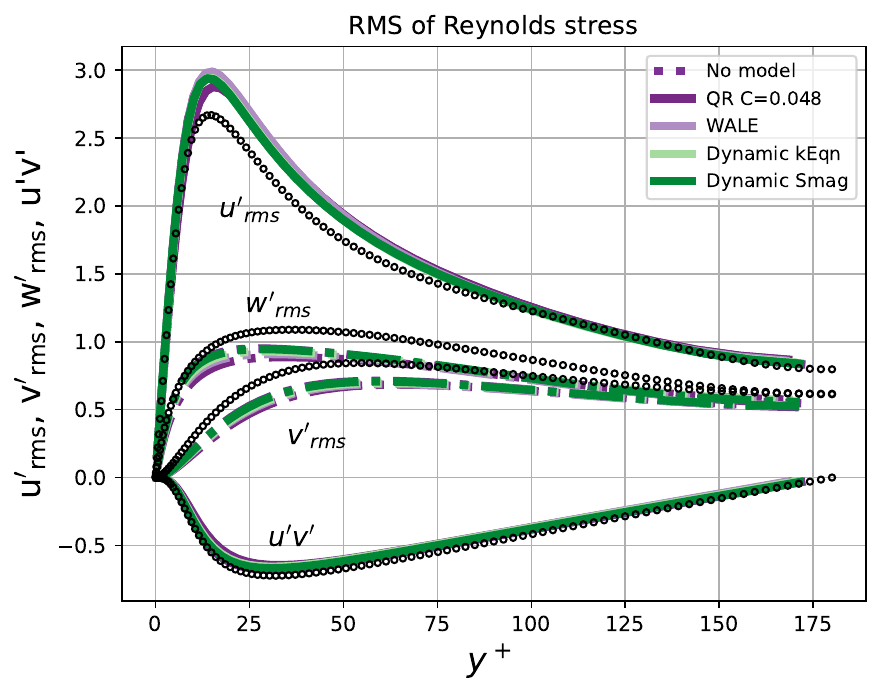}  
	\includegraphics[width=0.32\linewidth]{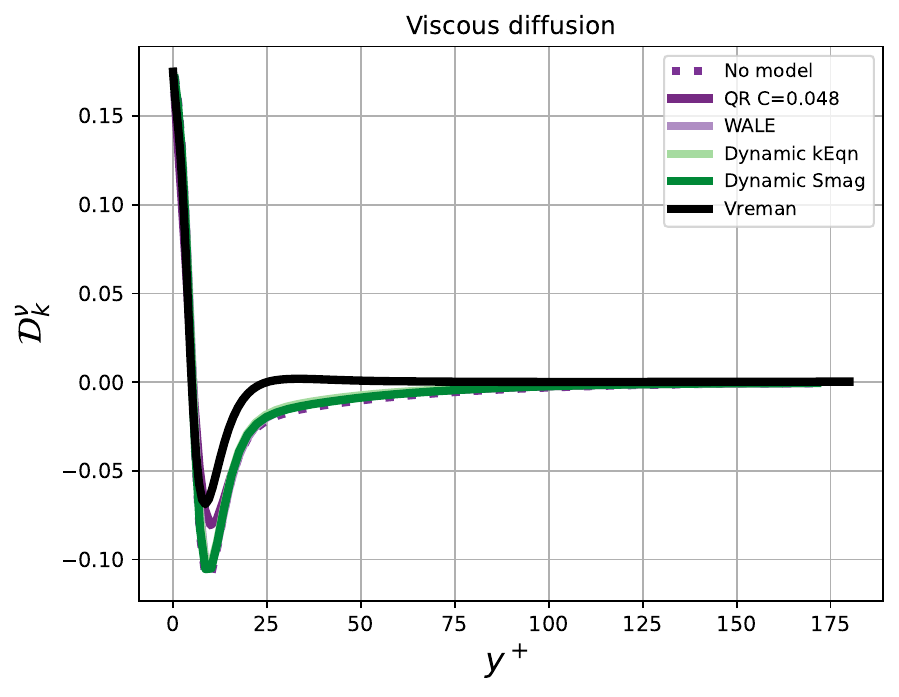}
	\includegraphics[width=0.32\linewidth]{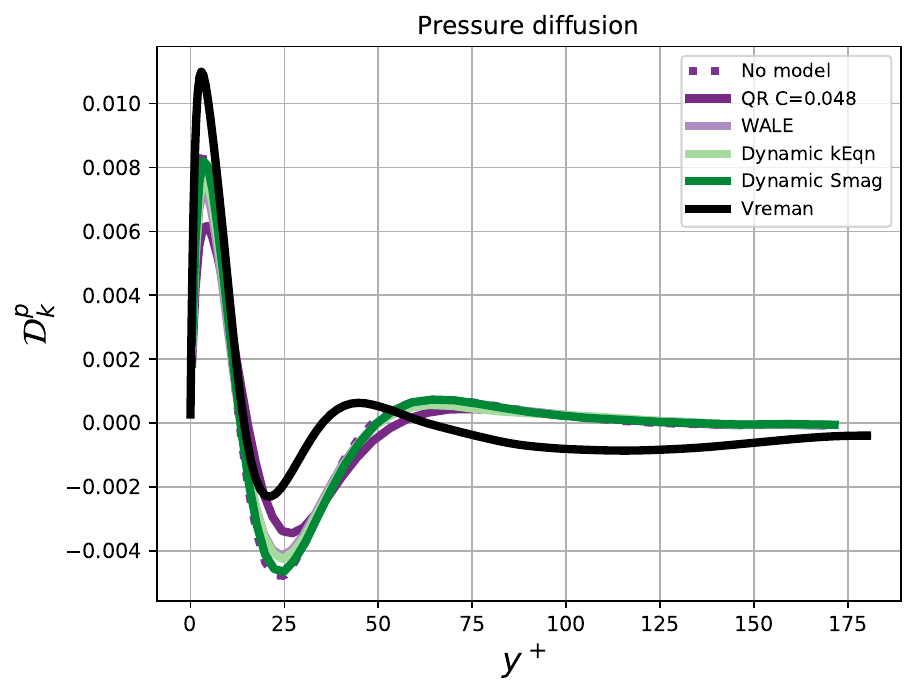} 
	\includegraphics[width=0.32\linewidth]{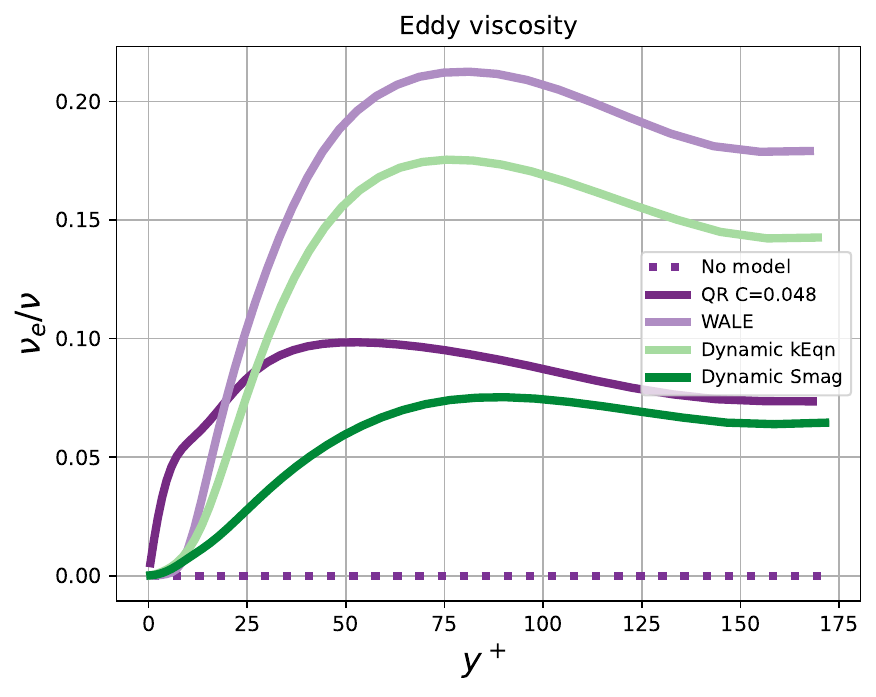}  
	\includegraphics[width=0.32\linewidth]{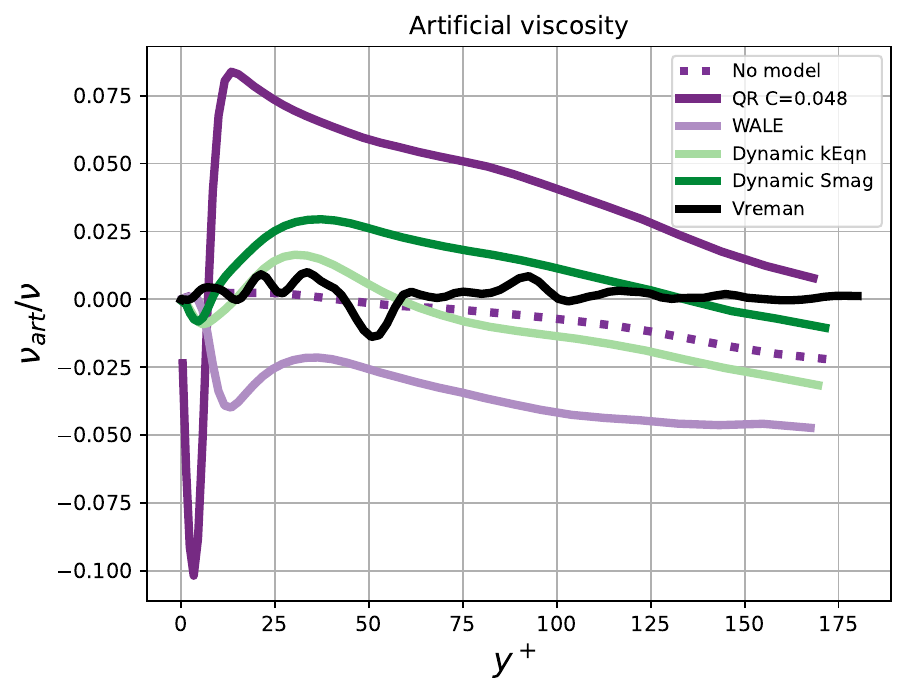} 
	\captionsetup{font = {footnotesize}}
	\caption{Mean and RMS velocity, turbulent kinetic energy budgets, artificial viscosity, and eddy viscosity versus wall-normal distance in channel flow simulations at $\mathrm{Re}_\tau = 180$, comparing various SGS models using symmetry-preserving discretization. }
	\label{fig:sgs1}
\end{figure} 
It might be speculated that the low eddy viscosity introduced by the minimum-dissipation model stems from its derivation, aiming to minimize the introduction of eddy viscosity to damp kinetic energy. To dispel this notion, we compare the QR and AMD models with Smagorinsky \cite{smagorinsky1963general} (excluding the wall model), k-Equation \cite{Yoshizawa1986}, WALE \cite{nicoud1999subgrid}, dynamic Smagorinsky \cite{germano1991}, and dynamic k-Equation \cite{kim1995} models using symmetry-preserving discretization. Detailed numerical settings are listed in Table \ref{tab:numSetting} and the profiles of the flow quantities are shown in Figure \ref{fig:sgs1}.

Among the models analyzed, the dynamic Smagorinsky, dynamic k-equation, and WALE models produced results comparable in accuracy to the QR and AMD models. These five models maintained relatively low eddy viscosity, with $\frac{\nu_e}{\nu} \leq 25\%$. The amount of eddy viscosity ($\nu_e$) generated by the WALE and dynamic k-Equation models is approximately twice that produced by the QR and dynamic Smagorinsky models. 
The distribution of eddy viscosity was nearly uniform across the wall-normal direction for most models, with the AMD model being an exception due to its concentration of viscosity in the near-wall region. 
The artificial viscosity introduced by these models varies significantly. The AMD model exhibited the highest artificial viscosity, followed by the QR model with a larger model coefficient of 0.048. Interestingly, only the WALE model exhibits entirely negative artificial viscosity, while the others introduce both negative and positive $\nu_{art}$ and display similar distributions. The QR model produces the highest amplitude of $\nu_{art}$.

The performance of SGS models at low Reynolds numbers or with very fine grid resolutions is often questioned due to the perceived low activity of the models in these regimes \cite{adrian2019}. However, our analysis challenges this viewpoint.
The eddy viscosity $\nu_e$ in Figure \ref{fig:sgs1} illustrates a nearly uniform distribution of eddy viscosity introduced by the SGS models in the outer wall layer. This uniform distribution signifies active damping of kinetic energy by the subgrid-scale models, even in low Reynolds number conditions. This finding suggests that SGS models remain actively engaged in energy dissipation across different flow regimes. 

\subsection{Optimizing the AMD Model Coefficient at Re$_\tau$ = 180}\label{sec:AMDcoeff}
\begin{figure}[b!]
	\centering 
	\includegraphics[width=0.32\linewidth]{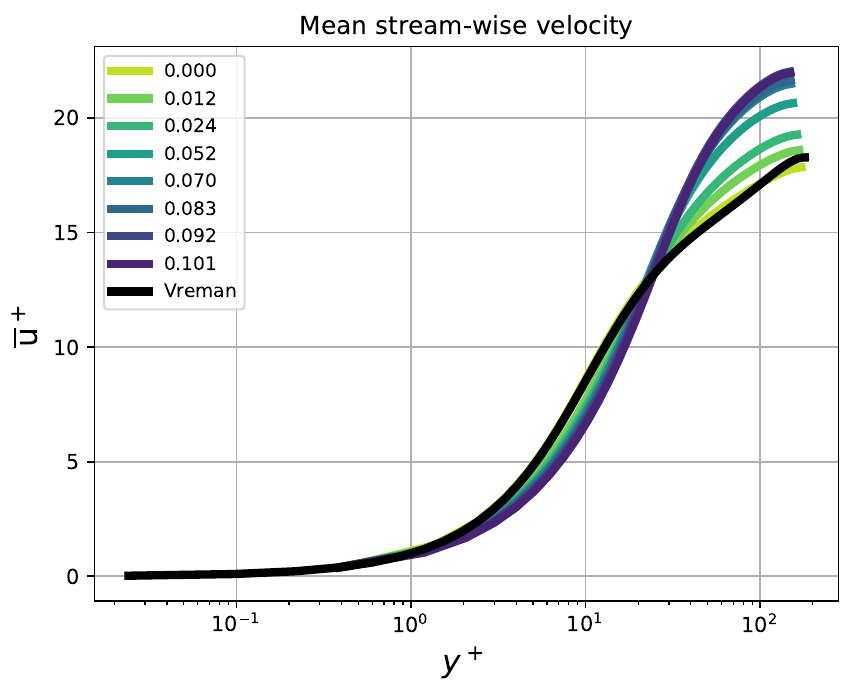}
	\includegraphics[width=0.32\linewidth]{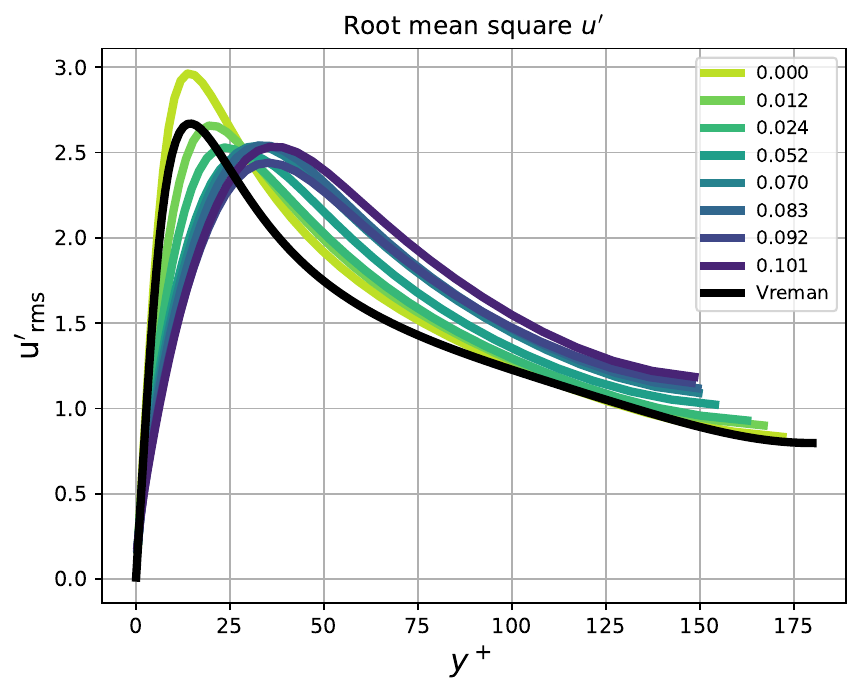} 
	\includegraphics[width=0.32\linewidth]{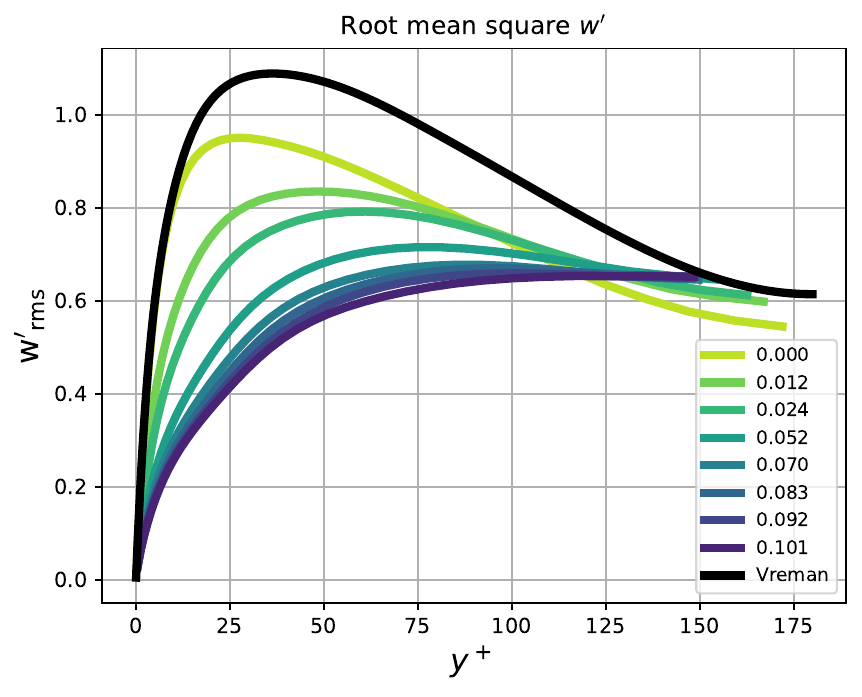} 
	\includegraphics[width=0.32\linewidth]{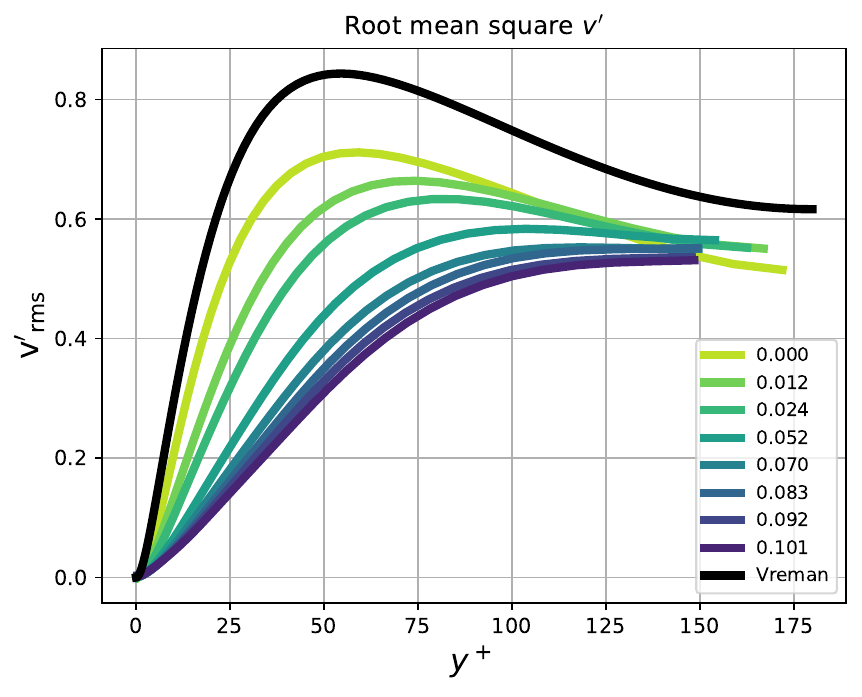} 
	\includegraphics[width=0.32\linewidth]{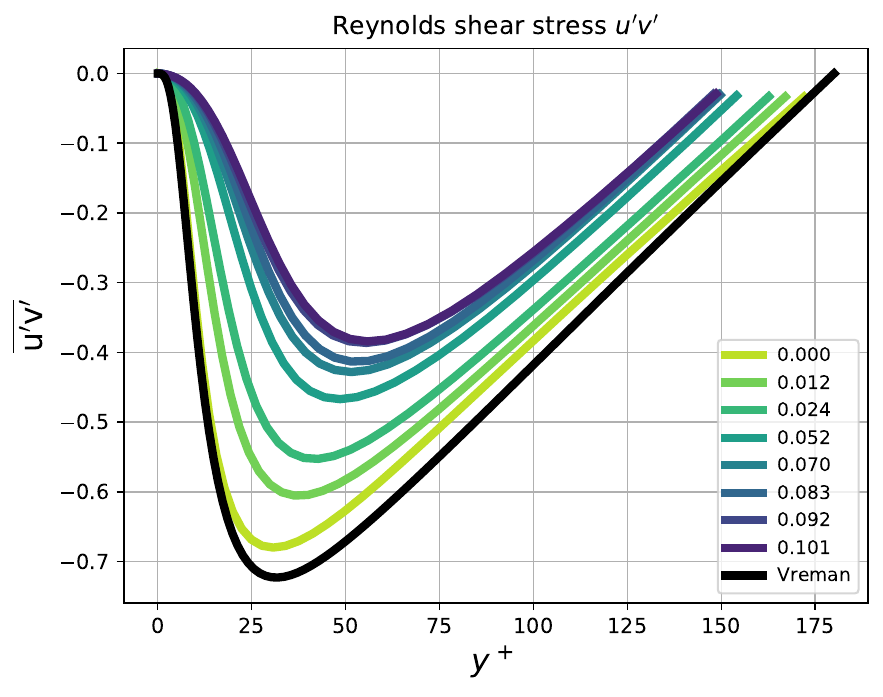} 
	\includegraphics[width=0.32\linewidth]{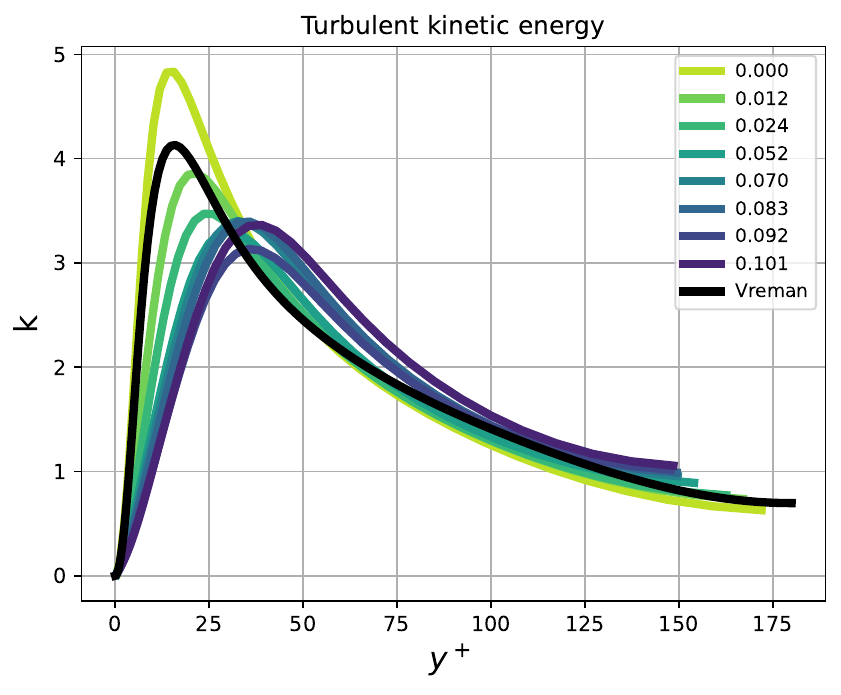} 
	\captionsetup{font = {footnotesize}}
	\caption{Mean, fluctuation velocity and turbulent kinetic energy versus wall-normal distance in channel flow simulations at $\mathrm{Re}_\tau=180$ for optimizing the AMD model coefficient using symmetry-preserving discretization. }
	\label{fig:AMDcoeff1}
\end{figure}
\begin{figure}[!t]
	\centering
	\includegraphics[width=0.32\linewidth]{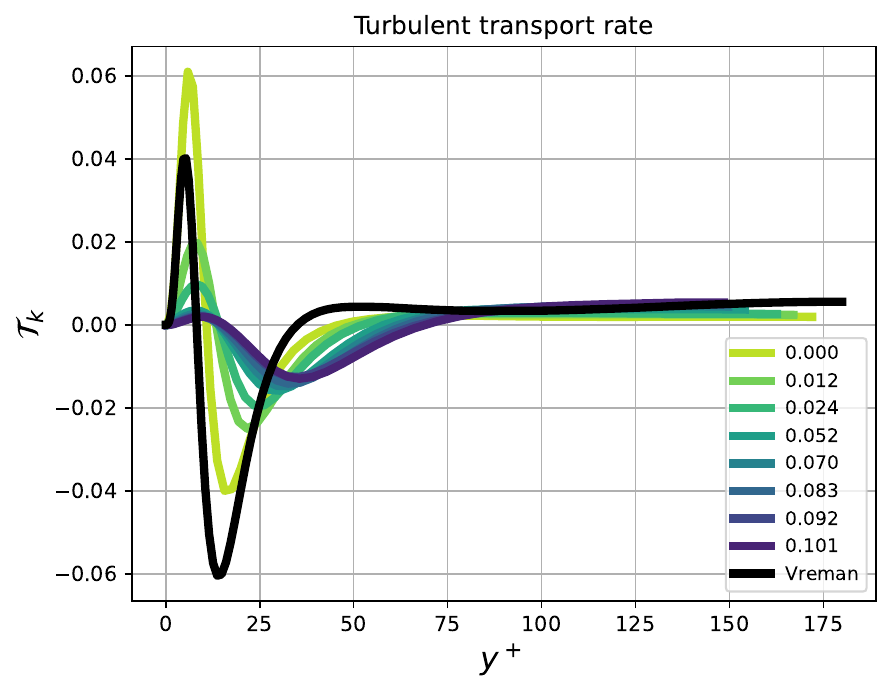}
	\includegraphics[width=0.32\linewidth]{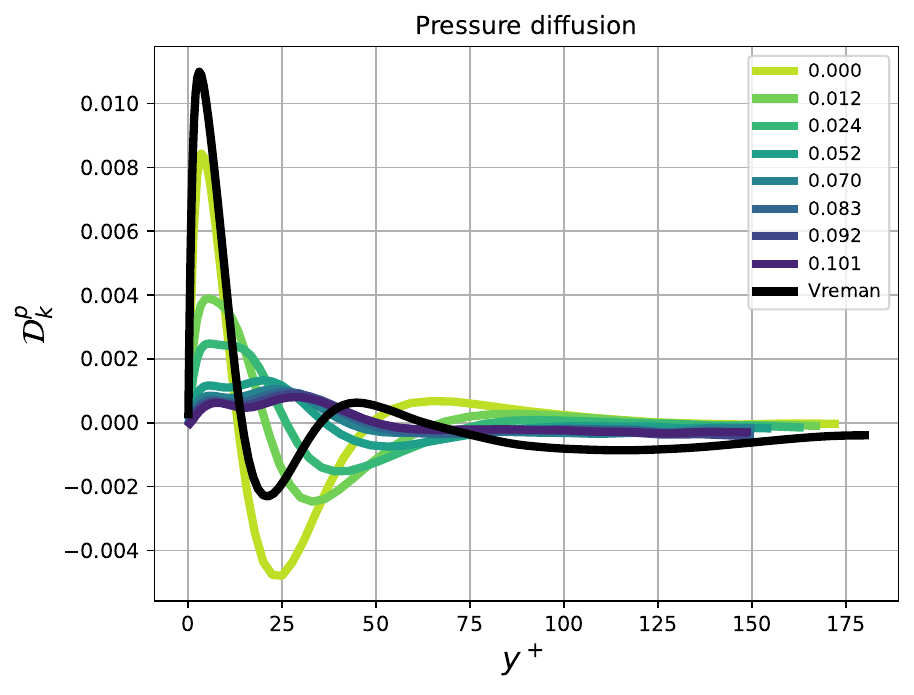}
	\includegraphics[width=0.32\linewidth]{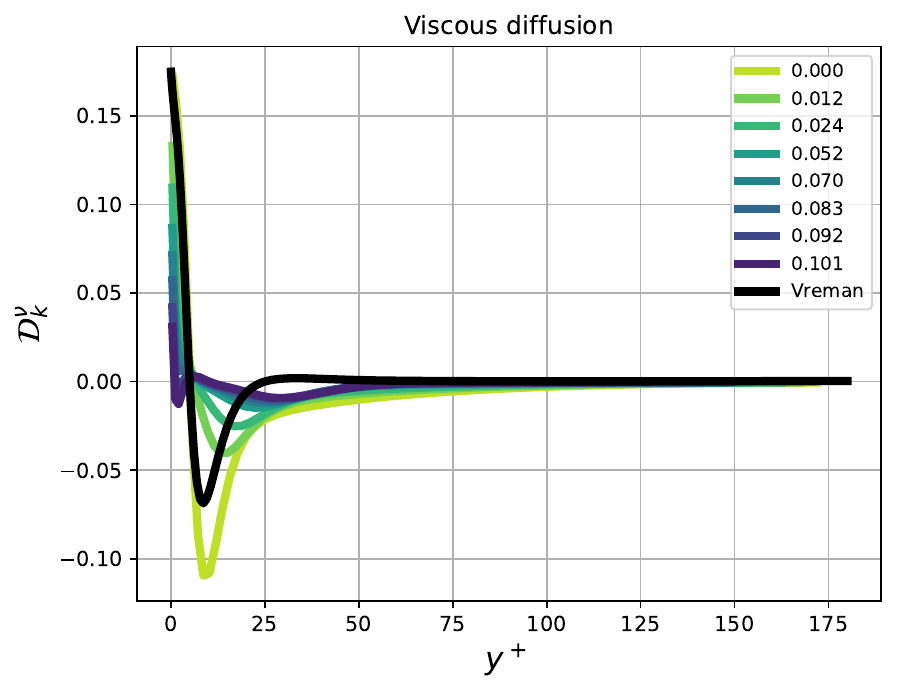}
	\includegraphics[width=0.32\linewidth]{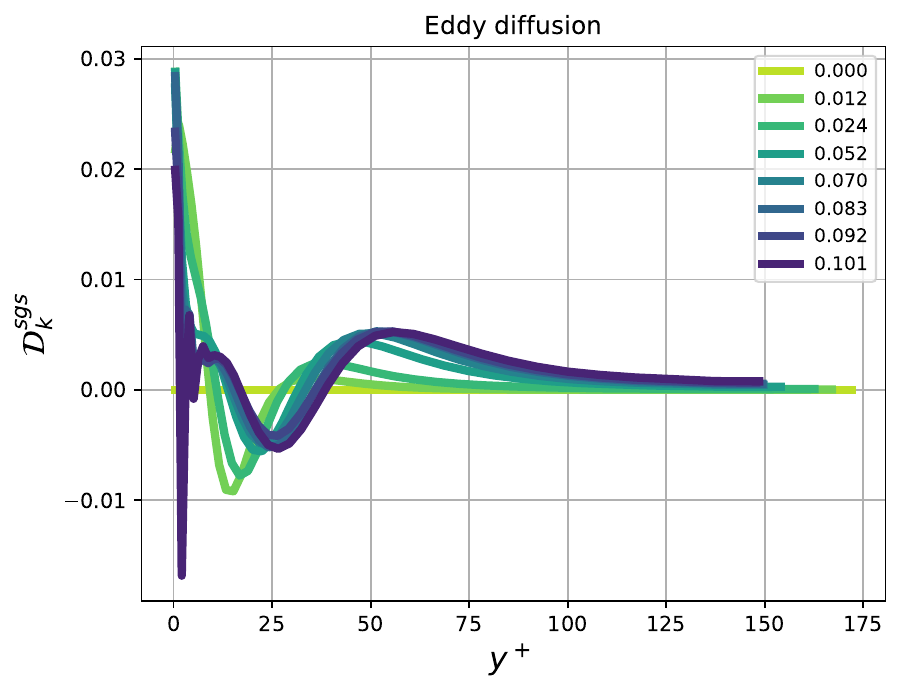}
	\includegraphics[width=0.32\linewidth]{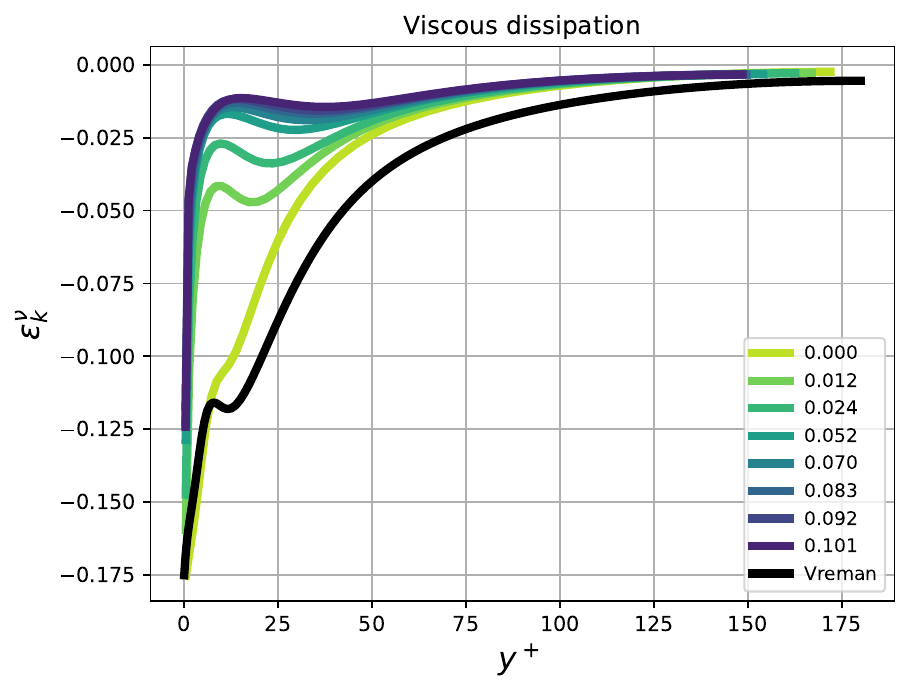}
	\includegraphics[width=0.32\linewidth]{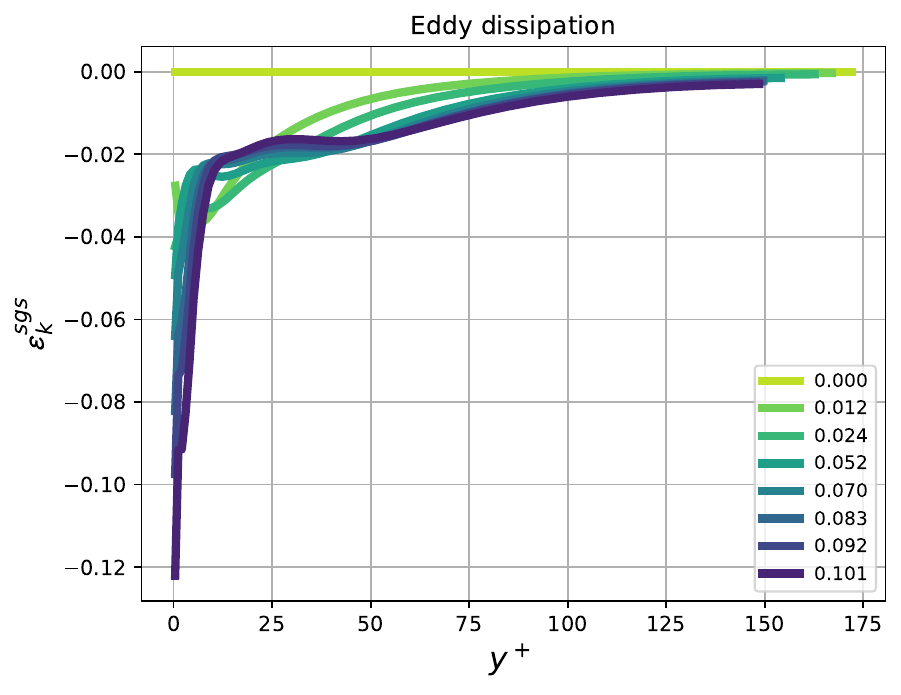}
	\includegraphics[width=0.32\linewidth]{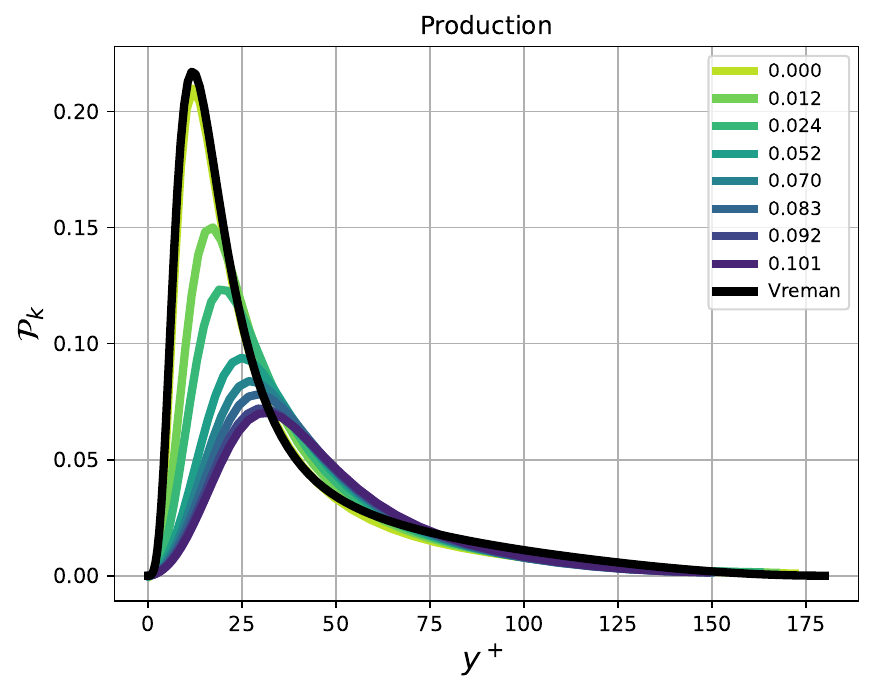}
	\includegraphics[width=0.32\linewidth]{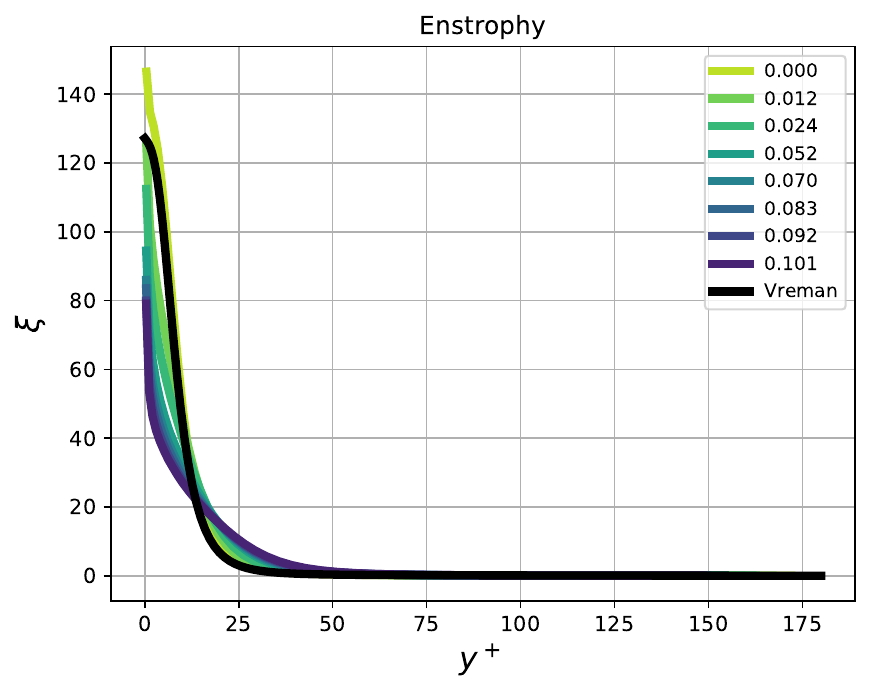}
	\includegraphics[width=0.32\linewidth]{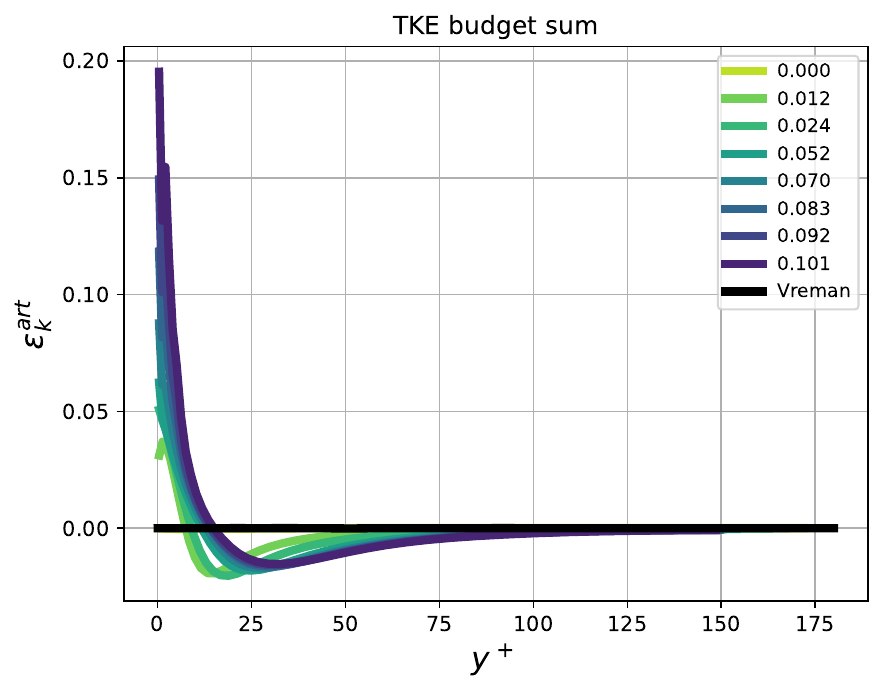}
	\captionsetup{font = {footnotesize}}
	\caption{Energy balance terms versus wall-normal distance in channel flow simulations at $\mathrm{Re}_\tau=180$ for optimizing the AMD model coefficient using symmetry-preserving discretization. }
	\label{fig:AMDcoeff2}
\end{figure}
\begin{figure}[!t]
	\centering
	\includegraphics[width=0.32\linewidth]{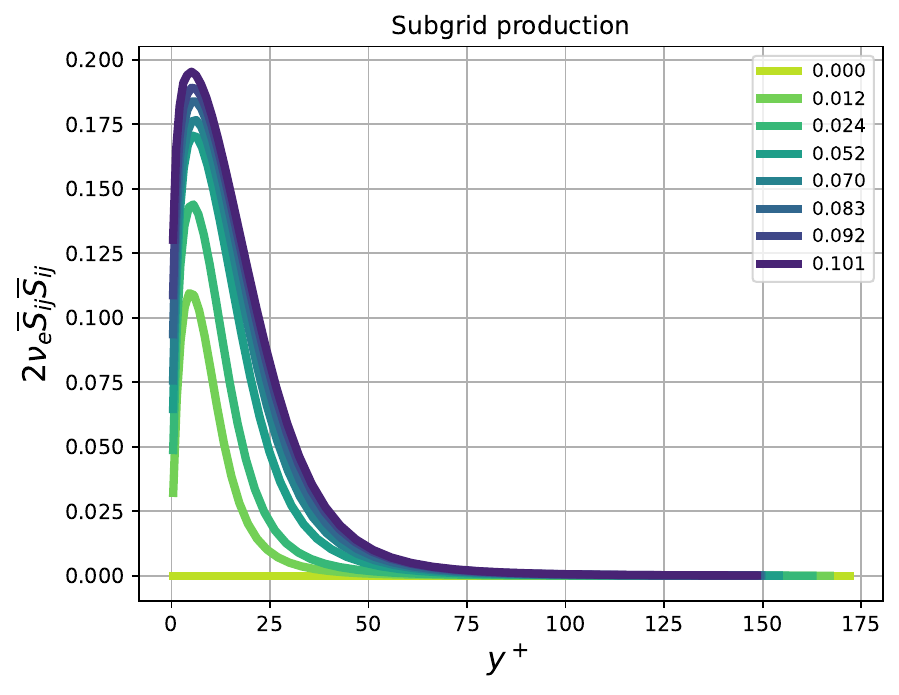}
	\includegraphics[width=0.32\linewidth]{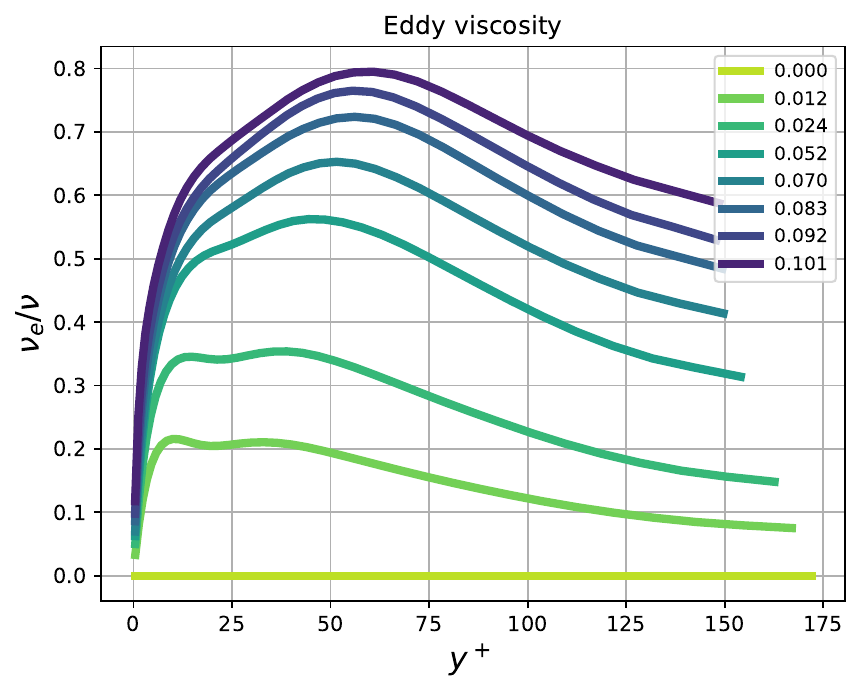}
	\includegraphics[width=0.32\linewidth]{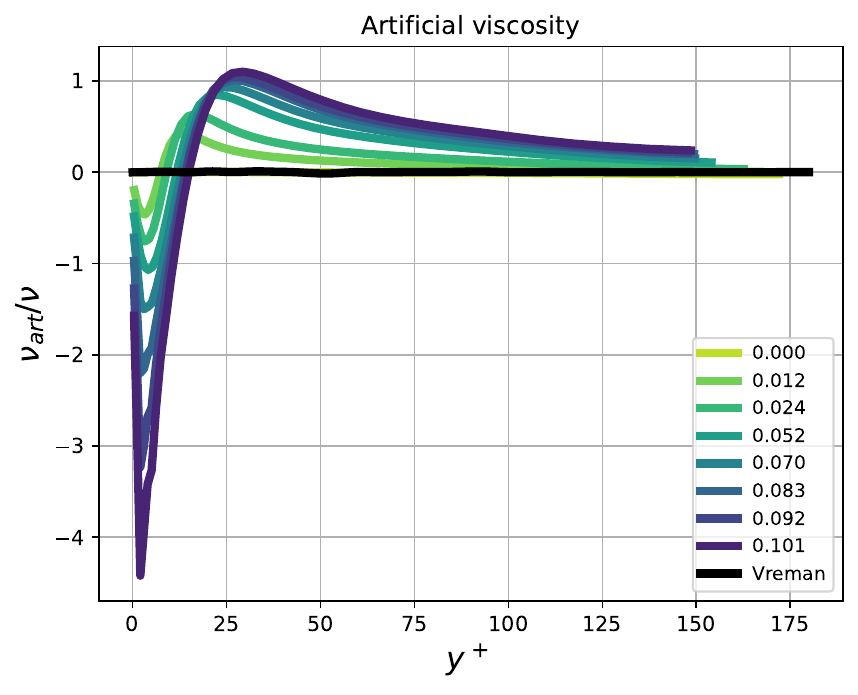}  
	\captionsetup{font = {footnotesize}}
	\caption{AMD model terms versus wall-normal distance in channel flow simulations at $\mathrm{Re}_\tau=180$ for optimizing the AMD model coefficient using symmetry-preserving discretization compared with DNS data.}
	\label{fig:AMDcoeff3}
\end{figure} 
\begin{figure}[t!]
	\centering 
	\includegraphics[width=0.42\linewidth]{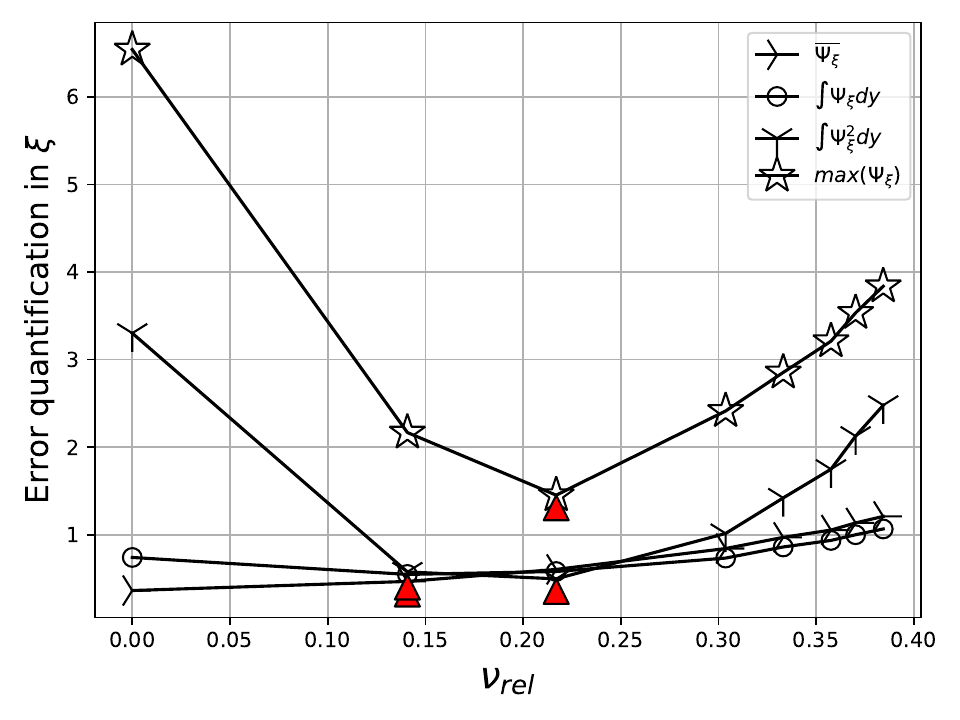} 
	\captionsetup{font = {footnotesize}}
	\caption{Error quantification in the kinetic energy of the mean flow $\bar E$,  shear stress $u'v'$ and turbulent kinetic energy $k$ in channel flow simulation at $\mathrm{Re}_\tau =180$ optimizing the AMD coefficient.}
	\label{fig:AMDcoeff4}
\end{figure} 
\begin{figure}[t!]
	\centering 
	\includegraphics[width=0.42\linewidth]{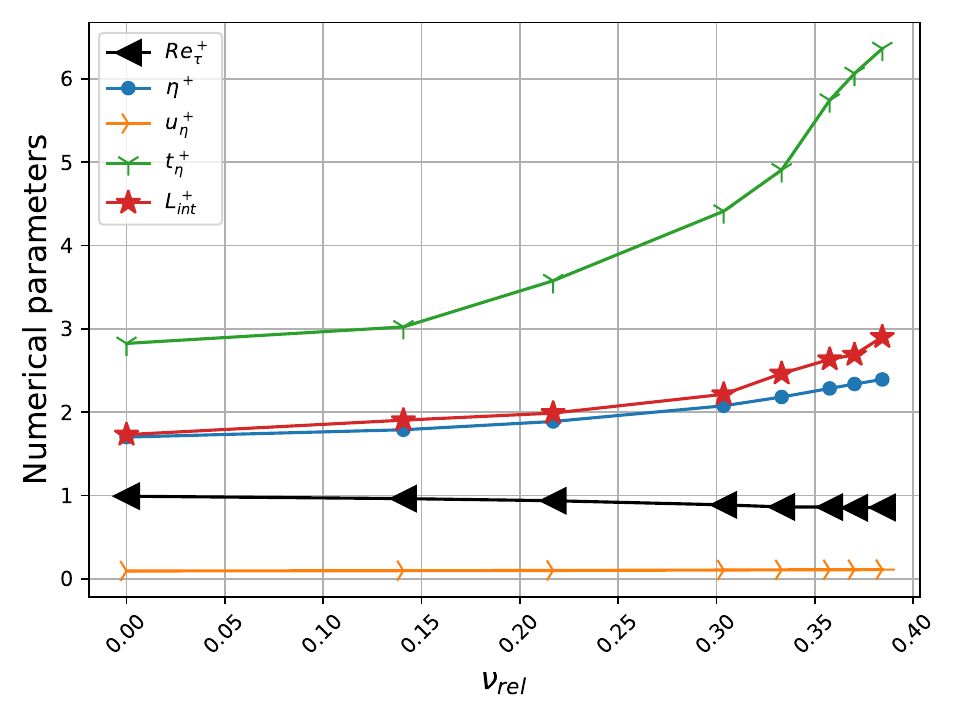} 
	\captionsetup{font = {footnotesize}}
	\caption{Numerical parameters versus the AMD coefficient in channel flow simulations  at $\mathrm{Re}_\tau =180$ optimizing the AMD model coefficient using symmetry-preserving discretization.}
	\label{fig:AMDcoeff5}
\end{figure} 
\begin{table}[h!]
\footnotesize
\centering
\begin{tabular}{ccccccccc}
\toprule
Coeff& $\mathcal P_k$ & $\epsilon^\nu_k$ & $\mathcal T_k$ & $\mathcal D^p_k$ & $\mathcal D^{sgs}_k$ &  $\mathcal D^\nu_k$ & $\mathcal P_k + \epsilon^\nu_k$ & $\epsilon^{sgs}_k$ \\ 
\midrule
0.00 & 0.0329 & -0.0254 & 9.5e-5 & -3.64e-6 &0.00& -0.0075 & 0.00743& 0.00  \\ 
0.012 & 0.0301 & -0.0173 & 4.39e-5 & -3.14e-6 &0.0004& -0.0047& 0.0128& -0.007 \\
0.024 & 0.0284 & -0.0146 & 2.89e-5 & -2.55e-6 & 0.0008& -0.0034&0.0138& -0.009 \\
0.052 & 0.0259 & -0.0120 & 1.19e-5 & -1.61e-6 & 0.0014& -0.0021&0.0139& -0.011 \\
0.083 & 0.0239 & -0.0104 & 4.20e-6 & -1.65e-6 & 0.0017& -0.0017&0.0135& -0.012\\ 
Vreman & 3.48e-2 & -3.48e-2 & -1.0e-12 & -1.72e-8 &0.00& -1.56e-5 & 2.44e-5&0.000 \\
\bottomrule
\end{tabular}
\captionsetup{font={footnotesize}}
\caption{Integrated turbulent kinetic energy budget over half of the channel height in fully developed channel flow at $\mathrm{Re}_\tau = 180$, investigating the optimization of the AMD model coefficient with symmetry-preserving discretization.}
\label{tab:AMDcoeff4}
\end{table} 
This section primarily aims to determine the optimal model coefficient for the AMD model Eq.\eqref{eq:AMD}. The coefficient values span from $C=0.012$ to $0.471$, with those above 0.083 derived from literature \cite{verstappen2011does,verstappen2018much,Rozema2020LowDissipationSM,rozema2015minimum}, while the others are chosen for consistency with the Smagorinsky model and practical considerations. Detailed numerical configurations are outlined in Table \ref{tab:numSetting}.

Figure \ref{fig:AMDcoeff1} presents profiles of mean velocity, RMS of Reynolds stress, kinetic energy, Figure \ref{fig:AMDcoeff2} shows subgrid kinetic energy balance terms, and Figure \ref{fig:AMDcoeff4} presents the error quantification results. The resulting physical parameters are summarized in Figure \ref{fig:AMDcoeff5}. Key trends include: \\
The AMD coefficient $C=0.012$ demonstrate minimal errors in the enstrophy $\xi$, the kinetic energy of the mean flow $\bar E$ and $u'v'$. As the model coefficient increases, Reynolds numbers decrease, Kolmogorov length and time scales increase, and boundary layers become thicker.\\
Regarding the RMS values of streamwise $u'u'$ , spanwise $w'w'$ and wall-normal $v'v'$ components of the Reynolds stress and the turbulent kinetic energy, the peak shifts away from the wall with the increase of the model coefficients. The change of the Reynolds stress in the near wall region ($y^+<25$) slows down. Larger under-predictions occur to the peak value of $w'w'$, $v'v'$ and $u'v'$, indicating reduced transfer of turbulent energy from streamwise to spanwise and wall-normal components.\\
For the subgrid kinetic energy balance terms, the exact Reynolds stress production rate $\mathcal P_k$ and the dissipation rate $\epsilon^\nu_k$ largely decrease. Meanwhile $\mathcal P_{sgs}$ increases dramatically and dominates kinetic energy production near the wall. Turbulent transport, pressure diffusion, and viscous diffusion rates of kinetic energy become less vigorous, approaching zero near the wall. Eddy viscosity is concentrated near walls and relatively small compared to molecular viscosity. No-model simulations exhibit zero budget sum, suggesting no global artificial dissipation. This net-zero artificial dissipation is achieved through the balance between the non-zero redistribution term and the non-zero sum of production and dissipation, as indicated in Table \ref{tab:AMDcoeff4}. 
 
\section{Conclusion} \label{sec:sum}
We have considered the error in LES, focusing on artificial dissipation, based on conservation of turbulent kinetic energy. This method was tested in LES of turbulent channel flow at $\mathrm{Re_\tau}=180$, using various subgrid-scale models. A particular focus is given to the minimum-dissipation model. The comprehensive comparison between our results and DNS results reveals several key insights:   
  
We found that the amount of artificial viscosity is comparable to the eddy viscosity, but their distributions are essentially opposite. The artificial viscosity can either produce turbulent kinetic energy (TKE) in the near wall region or dissipate TKE in the bulk region. The eddy viscosity is almost uniformly distributed across the wall-normal direction and actively damps TKE at the channel center. An eddy viscosity level of $\frac{\nu_e}{\nu} <22\%$ leads to accurate predictions of flow quantities at the current mesh resolution. 

Various minimum dissipation models on isotropic grids produce nearly identical results by introducing different amounts of eddy viscosity. Furthermore, these minimum-dissipation models outperform the no-model scenario.  
The optimal coefficients for the QR model combined with symmetry-preserving discretization were found to be $C=0.092$ for turbulent kinetic energy $k$ and $C=0.012$ for Reynolds stress terms and the kinetic energy of the mean flow $\bar E$. Adjusting the QR model coefficient changes the artificial viscosity and significantly impact turbulence characteristics: small-scale motions remain unaffected, while larger scales are less accurately captured with larger coefficients. Increasing the model coefficient beyond the optimal value leads to laminarization. The budget terms, except for $\epsilon^{sgs}_k$, exhibit self-similarity in the bulk region.  
For error quantification, the SGS activity parameter $\nu_{rel} = \frac{\nu_e}{\nu_e + \nu}$ helps relate and reflect the error magnitude with the SGS model.

Optimal coefficients for the AMD model were also identified. The trends of varying the coefficient are similar to the QR model but differ in the amplitude of changes. The AMD model is more sensitive to the model coefficients and introduces more eddy viscosity in the near-wall region compared to the QR model.

In conclusion, our quantification method provides a robust framework for understanding and harnessing numerical errors in LES. The interaction between numerical and eddy viscosities is crucial for accurate LES predictions. The QR model, with its optimal coefficients and energy conservation properties, stands out as a promising tool for future LES applications. 
\section*{Acknowledgements}
This work was supported by the Chinese Scholarship Council and the University of Groningen under Grant (CSC NO. 202006870021). The computing time was granted by the Dutch Research Council (NWO) under project 2022.009.

 
%
\section*{Data availability}
The used code is published on GitHub \cite{sunQR}. The numerical data for the channel flow can be made available upon request.
\bibliographystyle{elsarticle-num.bst} 
\bibliography{abbr_sources}
\end{document}